\documentclass[trackchanges]{aastex701}

\usepackage{comment}
\usepackage{gensymb}
\usepackage{hyperref}

\begin{document}

\title{Discovery of variable polarization in H$\alpha$ profile of symbiotic star Y Gem: A case for orbital-phase dependent variation of Raman-scattered Ly$\beta$ emission
}

\author[orcid=0000-0001-7340-8873,sname='Maiti']{Arijit Maiti}
\altaffiliation{Indian Institute of Technology Gandhinagar, Gandhinagar, India, 382055}
\affiliation{Physical Research Laboratory, Ahmedabad, India, 380009}
\email[show]{arijitmaiti@prl.res.in}  

\author[gname=Mudit K.,sname='Srivastava']{Mudit K. Srivastava} 
\affiliation{Physical Research Laboratory, Ahmedabad, India, 380009}
\email[show]{mudit@prl.res.in}

\author[gname=Vipin,sname=Kumar]{Vipin Kumar}
\affiliation{I. Physikalisches Institut, Universit at zu K\"oln, K\"oln, Germany, 50937}
\email{kumar@ph1.uni-koeln.de}


\begin{abstract}

The geometry and morphology of symbiotic stars are conducive to exhibit a variety of scattering phenomena. The prominent among them is the Raman scattering of O VI doublet $\lambda \lambda$ 1032,1038$\AA$, which often show strongly polarized features in the visible spectrum. Similar Raman scattering of Ly$\beta$ photons has also been predicted to occur in symbiotic stars, though with fewer detections and with weak polarization amplitudes. Here, we present the discovery of strong variable polarization in the H$\alpha$ profile of a recently established symbiotic system Y Gem, over a period of nearly 22 months. This is, most likely, a very rare detection of the strongly polarized Raman scattered Ly$\beta$ photons, falling at the H$\alpha$ emission. Monte-Carlo simulations have been conducted to confirm the underlying Raman scattering process causing the polarized line profile, and a simple orbital model is constructed with typical parameters available in the recent literature along with a complementary low-resolution spectroscopic data. These simulations and models are then used to validate the observed polarization variation of H$\alpha$ at different orbital phases corresponding to the epochs of observations. The possibility of such strong variable H$\alpha$ polarization, being caused by Raman scattering of Ly$\beta$, would thus open up avenues of exploring such effects in various other astrophysical situations having similar morphology.

\end{abstract}



\section{Introduction} \label{Intro}

Symbiotic stars (SySt) are long-period interacting binaries consisting of an evolved giant, either a red giant  (S-type) or a Mira variable embedded in a dust cocoon (D-type), and a hot companion, usually a white dwarf (WD) \citep{kenyon1986symbiotic}. This produces a distinctively composite stellar spectrum featuring strong emission lines superimposed on the spectral continuum consisting of broad molecular absorption bands \citep{kenyon1986symbiotic, van1993atlas, ivison1994atlas}. SySt generally have long orbital periods from a few years to decades. Their extended circumstellar environments, shaped by the giant’s mass loss and the ionizing radiation and winds of the hot star, host both ionized and neutral regions \citep{kenyon1986symbiotic, munari2019symbiotic, schmid1994raman}. Owing to this complex structure, SySt are key tracers of the late evolutionary phases of low- and intermediate-mass stars and provide a natural laboratory for studying binary interactions and evolution \citep{taylor1984radio, seaquist1984radio}. In recent times, significant work have been done to understand the underlying physical system with variety of observational data such as  time-domain photometry and spectroscopy from ground and space-based observatories \citep{munari2019symbiotic, merc2025symbiotic}, to further refine their classification criteria in order to understand their variable nature.
\par
 One of the most distinctive features of SySt is the presence of two broad features at $\lambda \lambda$ $6830, 7088 \text{\AA}$ in their optical spectra, which were noticed in a sample study of SySt \citep{allen1980unidentified}. Later, they were recognized as the Raman scattered features \citep{schmid1989identification,nussbaumer1989raman} of O VI $\lambda \lambda$ $1032, 1038 \text{\AA}$ resonance doublet. While the doublet emission originates from the hotter regions close to the WD, these photons get Raman scattered in the neutral hydrogen wind of the red giant, giving rise to such features in the visible spectrum. They became the signature identification features for the symbiotic stars. Nearly 50$\%$ of the symbiotics show these features \citep{van1993atlas, ivison1994atlas}. In addition to Raman scattering, the symbiotic phenomenon and its morphology are well suited for the scattering of different kinds (Rayleigh scattering, Thompson scattering, etc.) and for different atomic species \citep{schmid1994raman, ikeda2004polarized, lee2003raman, lee2012raman}. Another prominent one is the Raman scattering of Lyman $\beta$ (Ly$\beta$) from the neutral hydrogen winds of the red giant. Raman scattering of Ly$\beta$ photons is an inelastic radiative process wherein an incident Ly$\beta$ photon ($\sim$1025.7 $\AA$) excites a hydrogen atom from the ground state to a virtual intermediate level, after which the atom decays into the metastable 2s state rather than returning directly to the ground state. The excess energy is emitted as a longer-wavelength optical photon (around H$\alpha$) \citep{lee2000raman, yoo2002polarization, ikeda2004polarized}. While Raman-scattered features of O VI are mostly seen in symbiotics only, the Raman-scattering of Ly$\beta$ is predicted \citep{chang2015formation, arrieta2003broad, ikeda2004polarized} in several other astrophysical situations as well, such as symbiotic stars, planetary nebulae, active galactic nuclei, etc. It is worth noting that these Raman features are much broader than ordinary recombination lines, as the Raman scattering magnifies the dispersion width of incoming radiation by the ratio of the wavelengths of outgoing and incoming photons \citep{yoo2002polarization, arrieta2003broad, chang2015formation}.
\par 
Raman scattering of such emission lines in ultraviolet (UV) is a well-documented phenomenon in SySt, and as expected, strong polarization are observed in the Raman scattered $\lambda \lambda$ $6830, 7088 \text{\AA}$ features \citep{schmid1994raman, harries1996raman}. Polarization signatures were also reported in the broad H$\alpha$ wings of SySt, which is suggested to be caused by the Raman scattered Ly$\beta$ photons \citep{ikeda2004polarized, chang2018broad}. Symbiotic environments are rich to observe variety of other scattering processes. For example, Thomson scattering of the H$\alpha$ photons by the electrons would also lead to broadening of the H$\alpha$ line wings with a polarization signature \citep{kim2007Thompson, chang2018broad}. Polarized H$\alpha$ wings due to Thomson scattering have also been suspected in spectro-polarimetric observations of symbiotic star BI Crucis \citep{harries1996accretion}. For these reasons, the polarization studies of Symbiotics and their temporal variations are of great significance, as any change in the orbital geometry of SySt is expected to cause some variation in their polarization signatures in the spectrum. With the advent of surveys like GAIA and LAMOST, new SySt and SySt candidates are getting recognized \citep{ball2025symbiotic, akras2026discovery, chen2025new, zhao2025new}. Spectro-polarimetry of Raman-scattered features in their optical spectra can be an efficient tool to confirm their symbiotic origin. Furthermore, there is a potentially large population of non-shell burning SySt which hardly show any emission lines in their optical spectra but, their symbiotic nature has been proved by inspection of their ultraviolet (UV) spectra, as is the case for symbiotic star SU Lyn \citep{mukai2016lyncis, kumar2020uv, ilkiewicz2022lyn}.
\par
Y Geminorum (Y Gem) is one such system that was long classified as an AGB star with a main-sequence companion \citep{sahai2008binarity, sahai2011strong}. The nature of the companion, however, was long debated. Based on the UV excess, relatively low outflow velocity, and absence of narrow optical emission lines typical of symbiotic stars (e.g., H I, He II etc.), the companion was argued to be a main-sequence star \citep{sahai2018binarity}. In contrast, X-ray and far-UV observations and reflection modeling \citep{yu2022gem} suggested Y Gem to be a symbiotic star, where a white dwarf in a wide orbit accretes material from the AGB companion via wind Roche-lobe overflow. However, most recently, a multi-wavelength analysis \citep{guerrero2025gem} confirmed Y Gem’s symbiotic nature in favor of a WD. 
\par 
Here we present three H$\alpha$ polarization observations of Y Gem with a recently developed spectro-polarimeter named ProtoPol, over a period of nearly 2 years, viz., in March 2024, March 2025, and December 2025. While in the first epoch the H$\alpha$ profile did not show any significant polarization above the neighboring continuum, a distinctive polarization signature is recorded across the H-$\alpha$ emission in the second epoch, which further strengthened in the third epoch. Detection of strong H$\alpha$ polarization is rare in SySt, unlike their Raman scattered $\lambda \lambda$ $6830, 7088 \text{\AA}$ counterparts, though weak polarization has been reported in the H$\alpha$ wings in some cases \citep{ikeda2004polarized}. We had followed Y Gem earlier too (few epochs) around 2019 till 2020, with another in-house developed low-resolution spectrograph MFOSC-P \citep{srivastava2021design} on PRL 1.2m telescope, Mt Abu, India, for its temporal evolution. These observations are also presented here. The probable causes of the hitherto unknown detection of H$\alpha$ polarization in Y Gem are investigated with a detailed Monte-Carlo simulation of the Raman scattering process by assuming a simple orbital model of Y Gem constructed with typical parameters available in recent literature. The simulated profiles are compared to the observed ones, which suggest that the changes seen in the observed polarization profiles of H$\alpha$ are most likely related to the orbital phase of the Y Gem binary system.


\section{Observations and Data-Reduction} 
\label{subsec-OptObs}

Optical spectroscopic observations were conducted using the Mount Abu Faint Object Spectrograph and Camera–Pathfinder (MFOSC-P) instrument \cite{srivastava2021design} mounted on the PRL 1.2m telescope at the Mt. Abu Observatory, Gurushikhar (24.653 N, 72.779 E, 1680 m above mean sea level). MFOSC-P enables both imaging and spectroscopy within the same optical setup, with a wavelength coverage of 4500–8500 $\text{\AA}$. A log of the spectroscopic of Y Gem is presented in Table \ref{Table:Observation_Log}. The instrument delivers seeing-limited imaging in standard Bessel B, V, R, I, and H-$\alpha$ filters and is equipped with three plane reflection gratings - named R2000, R1000, and R500 - corresponding to their spectral resolutions of R = $\lambda/\delta\lambda$ = 2000, 1000, and 500,  having 500, 300, and 150 line-pairs per mm respectively. The standard wavelength coverages for these modes are 6000–7000 $\text{\AA}$ (R2000), 4700–6650 $\text{\AA}$ (R1000), and 4500–8500 $\text{\AA}$ (R500). The spectra and images are recorded on $1K \times 1K$ ANDOR make CCD detector. The instrument also contains a calibration unit, equipped with Neon, Xenon, and Halogen lamps, for wavelength calibration of the recorded spectra. These lamp spectra are usually collected immediately before or after the science observation frames. The raw data were processed using an in-house developed data-reduction pipeline built with custom Python routines based on open-source libraries, including ASTROPY. The reduction steps included bias and dark subtraction, cosmic-ray removal, spectral tracing and extraction, sky background subtraction, and wavelength calibration. Flat-fielding was not applied, as halogen lamp measurements showed pixel-to-pixel sensitivity variations below 1\% \cite{rajpurohit2020first, kumar2022optical}.
\par
Optical spectro-polarimetric observations were conducted using ProtoPol, a medium-resolution echelle spectro-polarimeter currently mounted on PRL 2.5m telescope at the Mt. Abu observatory, Gurushikhar \citep{kumar2022designs, srivastava2024development, srivastava2026development, maiti2026development}. ProtoPol covers the entire visible wavelength range from 4000-9600 $\AA$ with a spectral resolution of 0.4-0.75 $\AA$. A log of the spectro-polarimetric observations from ProtoPol is presented in Table~\ref{Table:Observation_Log}. The instrument couples a polarimeter unit to a spectrometer unit which employs echelle and cross-disperser (CD) gratings to produce cross-dispersed spectra of o- and e-ray for each echelle order on a $1K \times 1K$ ANDOR make CCD detector. Two separate CD gratings are used, one for the bluer wavelengths (4000-6000 $\AA$) and one for the redder wavelength regions (5800-9600 $\AA$). The instrument is also equipped with a calibration unit which houses a Uranium-Argon (UAr) lamp for wavelength calibration of the echelle orders and a halogen lamp for order tracing purposes. A full spectro-polarimetric data set consists of taking science exposures of the target star in 4 separate half-wave-plate (HWP) positions (0, 22.5, 45, and 67.5 degrees) along with calibration frames after each science exposure. The raw data from ProtoPol were processed using an in-house developed data-reduction pipeline built with custom Python routines based on open-source libraries like NUMPY, SCIPY, ASTROPY, etc. The data-reduction steps include bias and dark frame subtraction, cosmic-ray removal, order tracing, scattered background light subtraction, sky subtraction, extraction of o- and e- ray intensities for each order, wavelength calibration, and calculation of Stokes parameters from the extracted intensities of the orthogonally polarized rays.
The on-sky performance and characterization of the instrument, development and features of the data analysis pipeline, and a series of scientific observations of various kinds of astrophysical objects with ProtoPol are described in details in \cite{maiti2026development}.


\begin{table}[h]
\caption{Log of optical spectro-polarimetric and spectroscopic observations conducted with the ProtoPol and MFOSC-P instruments, respectively, on the PRL 2.5m and 1.2m Telescopes at the Mt. Abu Observatory, India.}\label{Table:Observation_Log}
\begin{tabular}{@{}ccccc@{}}
\hline
Observation date & JD & Instrument & Grating Mode (see text) & Exposure time (seconds)\\
\hline
14 November 2019 & 2458802.4 & MFOSC-P & R500, R1000, R2000 & 120s, 120s, 150s\\
07 December 2019 & 2458825.5 & MFOSC-P & R500, R1000, R2000 & 120s, 120s, 150s\\
04 January 2020 & 2458853.4 & MFOSC-P & R500, R1000, R2000 & 120s, 120s, 150s\\
20 March 2020 & 2458929.2 & MFOSC-P & R500, R1000, R2000 & 120s, 120s, 150s\\
25 December 2020 & 2459209.3 & MFOSC-P & R500, R1000, R2000 & 120s, 120s, 150s\\
06 March 2024 & 2460376.2 & ProtoPol & Red CD & 600s x 4 x 4 sets\\
01 March 2025 & 2460736.2 & ProtoPol & Red CD & 900s x 4 x 3 sets\\
16 December 2025 & 2461026.2 & ProtoPol & Red CD & 900s x 4 x 3 sets\\
\hline
\end{tabular}

\end{table}


\section{Analysis and results}
\label{sec-results}

\begin{figure*}
  \centering
  \includegraphics[width=\textwidth]{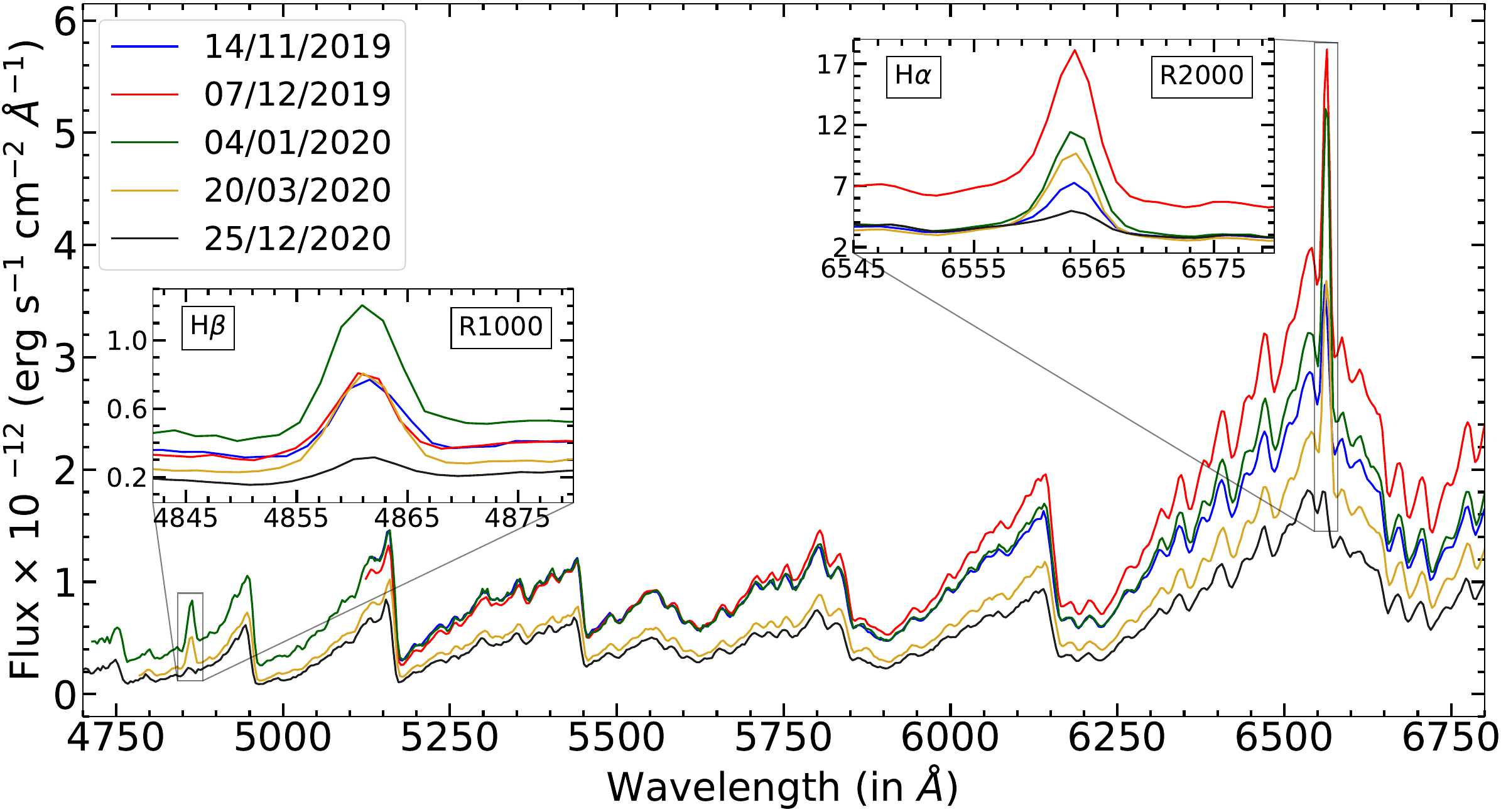}
  \caption{Low resolution ($\sim$500) optical spectra of Y Gem spanning 5 epochs from November 2019 to December 2020 (Table \ref{Table:Observation_Log}). The spectra have been corrected for extinction \citep{cardelli1989relationship} with $A_V = 0.8$ \citep{guerrero2025gem}. The flux-scaling of these spectra are done by pivoting them on the V-band magnitudes as obtained from AAVSO light curve. See text in section~\ref{sec-results}. The insets show the evolution of the H$\alpha$ and H$\beta$ emission profiles observed with resolutions of 2000 and 1000, respectively.}
  \label{Fig-MFOSC-P_Spectra}
\end{figure*}

Y Gem was first observed during the commissioning run of the then newly developed spectrograph MFOSC-P \citep{srivastava2021design} in 2019 and followed it for the low resolution (R$\sim500-2000$) optical spectroscopy for a year (see section~\ref{subsec-OptObs}). The spectra recorded in these epochs are shown in Figure~\ref{Fig-MFOSC-P_Spectra}. The relative flux calibration of the spectra was done using standard spectro-photometric stars observed on contemporaneous nights. We have pivoted the low-resolution spectra presented with reference to the V band magnitudes, as obtained from the AAVSO light curve database, by suitably interpolating the flux values of neighboring observations. The spectra were corrected for extinction \citep{cardelli1989relationship} with $A_V = 0.8$ \citep{guerrero2025gem}. During these epochs (spanning November 2019  - December 2020),  a significant change in the strength of the H$\alpha$ emission with respect to (w.r.t.) the neighboring continuum was noticed, and it got significantly weakened in December 2020. H$\beta$ also showed a similar trend (see insets of Figure~\ref{Fig-MFOSC-P_Spectra} for H$\alpha$ and H$\beta$ line evolution). The resolutions of these spectra (in the range of R$\sim$500-2000) were not sufficient to record any changes in the line profile variation. The continuum did not show any change in the slope of the spectral energy distribution (SED) over the wavelength range of the observations, which possibly indicates the minimal effect of the SED of the hot component on the SED of the cooler component of the Y Gem system. The MFOSC-P observations from November 2019 to March 2020 clearly showed an enhancement of the H$\alpha$ strength, w.r.t. neighboring continuum, reaching a maximum in March 2020. However, the next MFOSC-P observation in December 2020 showed a significant reduction in the H$\alpha$ strength to a level of almost complete disappearance. H$\alpha$ emission is known to be variable in several symbiotic systems due to a variety of reasons e.g., geometrical shielding effects, changes in activity of the emission region, etc \citep{mennickent2008remarkable, anderson1980observational, iijima1988high}. However, as discussed in later sections, these observations are used to construct a simple orbital model to explore the orbital phase dependence of the spectro-polarimetric profiles of Y Gem. 

\par 
The interest in Y Gem was again rekindled with the development of ProtoPol in 2023, and the system was again observed as a potential target for its spectro-polarimetry first in March 2024, followed by observations in March 2025 and December 2025. Between March 2024 and March 2025, the H$\alpha$ profile of Y Gem shifted from a single peak to an equal-strength double-peaked structure, which it retained in December 2025 too. The de-blending analysis of H$\alpha$ spectral profile reveals the presence of three major emission components: two narrow red and blue shifted emission, along with a broader underlying component (see Appendix~\ref{subsec-SpectralDeblending}). Over these three epochs, the relative strength of the double-peaked profile changed, with the red peak becoming stronger than the blue peak. The strength of the line with respect to (w.r.t.) the continuum also increased significantly.
\par
Apart from the hydrogen Balmer line emissions, the medium-resolution optical spectra from ProtoPol also reveal the presence emission lines of He I $\lambda$5876 $\AA$, forbidden lines like [N II] $\lambda$5755 $\AA$, [O I] $\lambda$6300 $\AA$ etc. However, unlike \cite{guerrero2025gem}, the [N II] $\lambda$6584 $\AA$ and [O I] $\lambda$6363 $\AA$ were a non-detection in ProtoPol data. The Raman scattered $\lambda \lambda$ 6830, 7088 $\AA$ features were also not detected in ProtoPol and MFOSC-P data, consistent with the results of \cite{guerrero2025gem}. The absence of any He I emission lines (except at $\lambda$5876 $\AA$) is indicative of low-ionization SySt. The optical spectrum of Y Gem is similar to other SySt hosting late red giants, for example, R Aqr and CH Cyg \citep{kafatos1991ultraviolet, burmeister2009spectroscopy}. Their spectra show the typical low-ionization lines like [Ne III] $\lambda$3869 $\AA$, [O III]  $\lambda$5007 $\AA$, [N II] $\lambda \lambda$5755,6584, [O I] $\lambda \lambda$6300,6363,  He I $\lambda$5876 $\AA$, etc. However, the Raman scattered $\lambda \lambda$ 6830, 7088 $\AA$ features, which are caused by the scattering of high ionization O VI $\lambda \lambda$ 1032, 1038 $\AA$, are also missing in those SySt. Similarly other high excitation lines like [Fe VII] $\lambda \lambda$ 5727, 6087 $\AA$ are missing. The non-detection of He II  $\lambda$4686 $\AA$ line in the typcial symbiotic spectra is an indication that the ionizing sources (WD and accretion disc) are not so hot (temperatures $<$ 37000 K) to provide high energy photons to ionize the high ionization potential species. \cite{sahai2018binarity} also suggested the temperature of the hot components in Y Gem to be in the range 35000-54000 K using SED fitting of the UV spectrum of the system.
\par

The above-discussed variation in the strengths and profile shape of H$\alpha$ is not a new phenomenon, and has been seen in various other symbiotics for varying reasons like geometric shielding effects \citep{iijima1988high, skopal2001photometric, lee2012analysis}, variation in flow velocity of circumstellar material \citep{anderson1980observational}, presence of hotspots in accretion discs \citep{robinson1994nature}, etc. However, it is the spectro-polarimetry of Y Gem - in particular of H$\alpha$ emission- which offers an entirely new perspective and insight into the orbital morphology of Y Gem system. These spectro-polarimetric observations of Y Gem over the three epochs from March 2024 to December 2025 reveal the detection of polarization in H$\alpha$ and its variation over the period of time. The strength of H$\alpha$ polarization is seen to be varied significantly from near-absence in March 2024 to around 4.5$\%$ (peak) in December 2025. To the best of our knowledge, the large amplitude polarization variation in H$\alpha$ profile has not been previously seen in any symbiotic system. Polarization in the H$\alpha$ wings has been observed in other symbiotic star systems like AG Dra and Z And \citep{ikeda2004polarized}, however, the strength of the polarization was found to be weak as compared to its Raman scattered 6830$\AA$ counterpart.
\par
\begin{figure*}
  \centering
  \includegraphics[width=\textwidth]{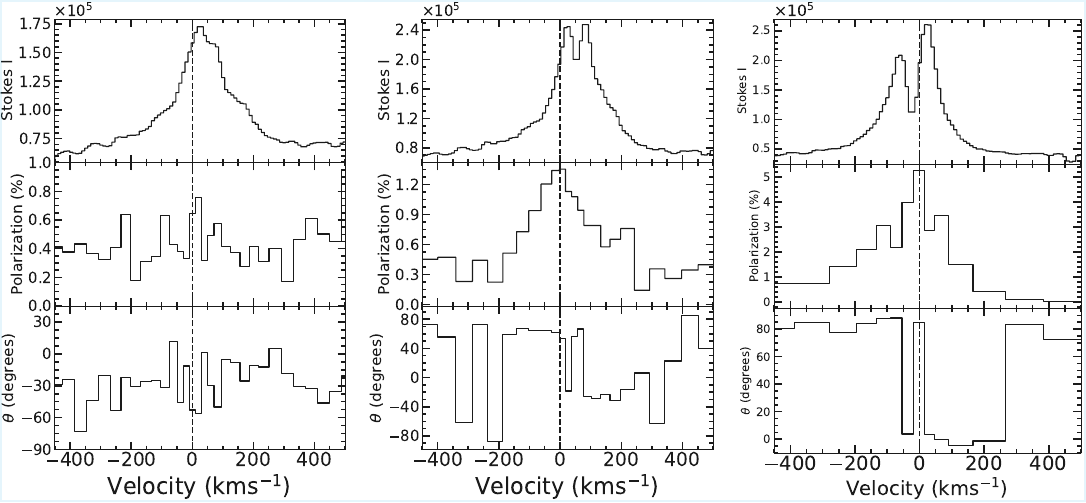}  
  \caption{Polarization profiles for H$\alpha$ as recorded in the spectra of symbiotic star Y Gem for March 2024 (left), March 2025 (middle), and December 2025 (right). Stokes parameter I (total intensity), the degree of polarization and the angle of polarization are shown in the top, middle, and bottom panel respectively. The data for the three observation dates have been binned to achieve a polarization accuracy of $0.15\%$, $0.12\%$, and $0.10\%$, respectively, per binning element.}
  \label{Fig-PolProfile}
\end{figure*}

 
The polarization profiles of H$\alpha$ in Y Gem for the said three epochs are shown in Figure~\ref{Fig-PolProfile}. No significant polarization was observed across the H$\alpha$ line in March 2024, though one could argue that around $\sim 0.3\%$ degree of polarization can be noticed at the peak of the profile above continuum just within 2$\sigma$ error. This observation was conducted during the commissioning test of ProtoPol instrument. The first detection of polarization in H$\alpha$ was rather serendipitous in the observations taken a year later (March 2025), however it was analyzed much later to first discover the signature of polarization in H$\alpha$. An immediate third observation in December 2025 was conducted, which revealed that the polarization profile got stronger (4.5$\%$ at peak). The angle of polarization showed almost 90$\degree$ flip across the H$\alpha$ profile for the last two epochs, similar to the one which had been modeled, predicted, and observed in the case of O VI Raman features \citep{schmid1994raman, harries1996raman, harries1997raman, lee1997profiles} in several symbiotic stars. The presence of multiple scattering regions and mirror symmetry of the scattering media with respect to the orbital plane of the symbiotic system were suggested to be the cause of the observed 90$\degree$ flip. However, it is most likely the first instance that such a flip is being observed in H$\alpha$. A rotation of the angle of polarization was also observed, similar to the angle of polarization rotation with the orbital motion of the system, reported in the Raman scattered O VI features in a few symbiotic systems like AG Dra, V1016 Cyg, SY Mus, etc \citep{schmid1997spectropolarimetry, schild1996spectropolarimetry, harries1996spectropolarimetric}.

\par 
Phase locked polarization angle rotation or changes in degree of polarization have been reported before for symbiotic star systems \citep{schild1996spectropolarimetry, schmid1997spectropolarimetry}, but only in the Raman scattered $\lambda \lambda$ $6830, 7088 \text{\AA}$ lines. Although H$\alpha$ polarization has been reported in very few cases for symbiotic stars, no reports of phase-locked polarization changes across the line have been reported. Furthermore, the reported polarization values across the H$\alpha$ line are usually a lot weaker than what is observed for Y Gem. This raises the question as to why such polarization and its changes have not been observed before in symbiotic systems. If the polarization is indeed caused by Raman scattering of Ly$\beta$ photons \citep{lee2000raman, ikeda2004polarized}, one reason could be the dilution of the polarized photons by the actual H$\alpha$ photons, resulting in weak/no observed polarization signature. Phase-locked polarization changes have been reported for other binary star systems as well \citep{brown1978polarisation, rudy1978polarimetric}, Wolf-Rayet stars \citep{st1993polarization}, etc., although these results were reported for imaging polarimetric data rather than polarization changes across a specific emission line.

\section{Discussion}

\subsection{Variability in H$\alpha$ polarization as a probe of orbital phase}
\label{subsec-Halpha-SpecPol}

Spectro-polarimetry of Raman scattered features of O VI $\lambda \lambda$ $1032, 1038 \text{\AA}$ in symbiotic star spectra show strong polarization signatures \citep{schmid1994raman, schild1996spectropolarimetry, harries1996raman, schmid1997spectropolarimetry}. Various modeling efforts of the Raman scattering phenomenon \citep{schmid1992montecarlo, schmid1995monte, schmid1996simulations, harries1997raman, lee1997profiles} have helped uncover the underlying physics and scattering morphology in the symbiotic star environment. However, an equally important scattering phenomenon is the Raman scattering of Ly$\beta$ photons in far UV by neutral Hydrogen (HI) wind of the red giant star, which results in the formation of broad H$\alpha$ wings caused by the Raman scattered photons \citep{lee2000raman, yoo2002polarization}. Many SySt are known to show broad wings in the H$\alpha$ profile \citep{lee2000raman, chang2018broad}, where the hydrodynamical motion of the system is also suggested to be a possible reason. In such a scenario, spectro-polarimetry has emerged as a tiebreaker technique, as if Raman scattering is the cause of broader wings, a polarization is expected to be detected in such broad wings, which indeed has been the case in some symbiotic stars like AG Dra and Z And \citep{ikeda2004polarized}. 
\par
The spectro-polarimetric observations presented in section~\ref{sec-results} for the three epochs spanning nearly 22 months (March 2024 - December 2025) depicted a rather unique dataset of the detection $\&$ evolution of polarization across the H$\alpha$ profile over a period of time. While, as discussed above, the polarization in H$\alpha$ has been reported in a few symbiotics \citep{ikeda2004polarized}, the detections were mostly confined to the wings only with polarization amplitudes of $\sim$0.5-0.6$\%$. It is only the O VI Raman scattered features (at $\lambda \lambda$ $6830, 7088 \text{\AA}$) where the broad polarization profiles with a large value of line polarization (even greater than 10$\%$) had been reported \citep{schmid1994raman, harries1996raman}. However, the polarization profiles of H$\alpha$ in Y Gem, shown in Figure~\ref{Fig-PolProfile}, revealed a polarization amplitude similar to that seen in the O VI Raman scattered features. The strong polarization, the unmistakable flip in the angle of polarization across the profile, and the phase-locked variation in the magnitude of line polarization, although reported in the Raman scattered O VI $\lambda \lambda$ $1032, 1038 \text{\AA}$ lines of some SySt, have not been previously reported in the H$\alpha$ line in symbiotic stars, to the best of our knowledge. If such is the case, the present dataset and the deduction discussed below could possibly be of significance not only for the studies of symbiotics but also for various other similar types of astrophysical systems, such as planetary nebulae, young stellar objects, etc. \citep{schmidt1981spectropolarimetry, oudmaijer1999halpha}.
\par
Raman scattering of Ly$\beta$ photons by neutral hydrogen remains the most plausible cause to explain the observed polarization profile, though elastic Thomson scattering of H$\alpha$ photons by fast-moving electrons has also been suggested to cause intrinsic polarization in H$\alpha$ in at least one symbiotic star \citep{harries1996accretion} and several other objects like Herbig stars, classical Be stars, binary systems, etc. \citep{harrington2007spectropolarimetry, oudmaijer1999halpha, brown1978polarisation, rudy1978polarimetric}. The broad wings of H$\alpha$ profile could also be formed by Thompson scattering, wherein the H$\alpha$ photons get scattered off fast-moving electrons, thus broadening the wings \citep{kim2007Thompson}, \citep{chang2018broad}. The Thompson scattering optical depth and electron temperature of symbiotic stars AG Dra, Z And, and V1016 Cyg have been determined by analyzing the broadening of wings of O VI and He II lines \citep{sekeravs2012electron}. However, it should be noted that there have been very few observational evidences that conclude electron-scattering of H$\alpha$ photons as the cause of the broader wings, e.g., electron scattering has been suggested as the cause of the polarized H$\alpha$ wings, from the spectro-polarimetric observations of the symbiotic star BI Crucis \citep{harries1996accretion}.
\par
While Thompson scattering could very well play a role in the detected polarization profile of H$\alpha$ in Y Gem, the similarity of the observed polarization amplitudes and phase variation of H$\alpha$ with that of the O VI Raman features led us to believe that Raman scattering of Ly$\beta$ could have played a significant role in the observed profiles. We, therefore, have constructed a simple model of the Y Gem system, based on various system parameters available in the literature, and performed a detailed Monte-Carlo simulation of the scattering process to determine the expected H$\alpha$ polarization profile as a function of the orbital phase. The model geometry, methodology, simulated results, etc., are described in detail in Appendix~\ref{subsec-MonteCarlo}.  The system parameters of Y Gem, such as masses of binary components, their effective temperatures, their separation, wind velocity, mass loss rate, etc., have been taken from recent literature \citep{yu2022gem, guerrero2025gem}. It should be noted that the aforementioned simple orbital model is constructed to validate the orbital phase dependence of the scattered-Ly$\beta$ induced polarization of H$\alpha$ and not to be considered as the orbital geometry of Y Gem system.


\subsection{A simple model of orbital phase for the observed Ly$\beta$ induced Polarization}
\label{subsec-OrbitalPhase}

We have simulated the phase-resolved polarization profiles (Appendix~\ref{subsec-MonteCarlo}) for Y Gem, which in turn allows us to construct an orbital phase model. Several system parameters (such as the orbital period, masses of the binary components, their separation, etc.) were obtained from a recent study \citep{guerrero2025gem}. In their analysis \cite{guerrero2025gem} considered the masses of the red giant and WD to be 1.1 M$_\odot$ and 0.8 M$_\odot$, respectively; the radius of the red giant to be 240 R$_\odot$; and the semi-major axis of the binary orbit was taken to be 5.3 AU.  However, one of the important derivation of their analysis, consequential to this study, was the suggestion of the orbital period of the Y Gem system. Through Lomb-Scargle periodogram analysis of the light curve of Y Gem (obtained from AAVSO\footnote{\href{https://www.aavso.org/}{https://www.aavso.org/}} database spanned over $\sim$90 years, \cite{guerrero2025gem} had identified six distinct periods (see section - 3.2 of their article) that were above the false alarm probability (FAP) level. Of those, the periods of 250 and 360 days were considered spurious, with the former arising from observational gaps and the latter due to the orbital period of Earth. The shortest detected period of 148.4 days was predicted due to pulsations of the long-period variable M-type star in Y Gem, while a group of periods, detected between 760-2570 days, were predicted to be caused by pulsations of long secondary period (LSP). The period of 8.87 years was the one suggested by \cite{guerrero2025gem} as the orbital period of the system, which is consistent with the estimate of \cite{yu2022gem} who inferred the orbital period of the WD companion in the Y Gem system between 8.7 and 12.9 years. Two other strong peaks were also detected in the periodogram at 28.9 and 65.9 years, but \cite{guerrero2025gem} notes that these periods could be resonance frequencies of the shorter 8.87-year period or they could be related to the magnetic activity cycle of the giant component of Y Gem. It is nevertheless noted that AAVSO data would not contain many cycles of these longer periods, making it difficult to test their reality. In this study, we presented that the Y Gem system showed variability, whether spectroscopic or spectro-polarimetric (section~\ref{subsec-OptObs}) over the time scale of a year or so. Therefore it seems more plausible to adopt the the shorter 8.87-year period  as the orbital period of Y Gem system. Though we do not claim it to be the true period of the system, it appears to be the most reasonable choice to build a simple toy-model of the Y Gem's binary orbit in order to explore the causes of spectro-polarimetric variation seen in the H$\alpha$ emission of Y Gem.
\par
We accept that, due to the lack of a sufficient number of data points, the ephemeris of the system or some of these parameters (such as orbital period, etc.) may not have been fully established beyond doubt. Typically it would be preferred to have sufficient large observational coverage spanning few orbital periods (or at least a large fraction of the orbital period) to make any deduction about the orbital period. Nevertheless, the small set of spectro-polarimetric observations presented here, as it would be shown, matches well with the predicted polarization trends from the Monte Carlo simulation. Thus, for the reasons discussed above we continue to build a proof-of-concept orbital-phase model (with aforementioned parameters) to demonstrate that the observed polarization in H$\alpha$ are most likely caused by the Raman scattering of Ly$\beta$ photons and its temporal variations are due to the morphological/geometrical changes occurring during the course of orbital-phase. We again reiterate that here we do not attempt to present the orbital ephemeris of Y Gem system, but to demonstrate that the Raman scattering of Ly$\beta$ photons is the mostly likely cause of the polarization seen in H$\alpha$.
\par
The relative phase positions of our observations on this orbital plane were determined from visual optimization of the eccentricity and rotation angle of the orbit to match the simulated polarization profiles with the observed ones. Here, we hasten to clarify that this may not be the unique orbital solution, as multiple solutions may be possible, for example, by choosing a different value of any free parameter, like eccentricity. Our aim here is just to demonstrate that the scattering of Ly$\beta$ phenomenon would satisfy the typical orbital geometry of a SySt and hence should be a highly probable cause of the observed polarization in H$\alpha$.
\par 
Though the observed variation in the strength of H$\alpha$ in Y Gem (see Figure~\ref{Fig-MFOSC-P_Spectra}) could be caused by any or a combination of reasons, as discussed in literature e.g., geometrical shielding effects, changes in activity of the emission region, etc. \citep{mennickent2008remarkable, sonith2023tcp, lee2012analysis, anderson1980observational, iijima1988high, kenyon1986symbiotic}, here we pivot the epoch of minimum of H$\alpha$ in December 2020 as the reference zero-point to fix the phase 0.0 of the orbital motion, wherein the hot WD and red giant are along the line of sight. The rest of the observing epochs (both before and after this zero-point) have been pivoted appropriately, as shown in Figure~\ref{Fig-phase_geometry}. The fixation of epochs of spectro-polarimetric observations, viz. March 2024, March 2025, and December 2025 on this orbital phase plot, then provided the inputs for the projected geometry of the binary system with respect to the observer for which the polarization profiles have been simulated and verified with the observed ones. Phase-locked variation in H$\alpha$ strength is a well-known phenomenon and has been discussed extensively in the literature \citep{anderson1980observational, iijima1988high, robinson1994nature, contini1997evolving, sonith2023tcp}. The above choice of the pivot epoch appears to be a good reference, as the diminishing of H$\alpha$, would require the system to be close to edge-on, which is consistent with the simulated polarization profiles (discussed below). Considering the radius of the red giant to be 240 $R_\odot$ and a semi-major axis of the orbit as 5.3 AU \citep{guerrero2025gem}, the expected inclination angle of such a system is $90^\circ \pm 13^\circ$. This model of orbital phase is similar to the one that has been reported for another edge-on symbiotic star system AX Per \citep{skopal2001photometric}, which also shows variation in the H$\alpha$ strength.
\par
The relative positions of the epochs of spectro-polarimetric observations are shown in the orbital phase plot (Figure~\ref{Fig-phase_geometry}). No clear polarization signature across H$\alpha$ was detected in the first spectro-polarimetric observations in March 2024, whereas a clear increase in polarization across the H$\alpha$ profile was evident from similar observations in March 2025 and December 2025. By adopting the aforementioned orbital phase for the Y Gem, the Monte-Carlo simulations of the Raman scattered Ly$\beta$ photons are used to construct the polarization profiles (degree and angle of polarization versus wavelength) at various phases/epochs and for various inclination angles (Appendix~\ref{subsec-MonteCarlo}). Generating models at various inclination angles suggested that there should be no change in the observed degree of polarization (w.r.t. phase), when the orbital plane is viewed face-on (inclination $\sim$0$\degree$), whereas maximum change is expected when the system is viewed edge-on (inclination $\sim$90$\degree$). The significant changes in the polarization seen in the H$\alpha$ profile w.r.t. orbital phase further strengthen the hypothesis that Y Gem could be a near edge-on system. Other recent studies of Y Gem \citep{yu2022gem, guerrero2025gem} also suggested that the Y Gem is an inclined system. The ratio of peak polarization between conjunction and phases closer to quadrature also indicates the system is expected to have a high inclination angle (Figure 15 of \cite{schmid1992montecarlo}). An additional support for high inclination angle comes from the 90$\degree$ flip observed in the polarization angle across the H$\alpha$ profile from Blue to Red wings. This is more prominent in the polarization profile of December 2025. Such 90$\degree$ flips in the angle of polarization have been predicted and observed in other high-inclination SySt such as HBV 475 and SY Mus \citep{schmid1994raman, harries1996spectropolarimetric}. 

\begin{figure*}
  \centering
  \includegraphics[width=\textwidth]{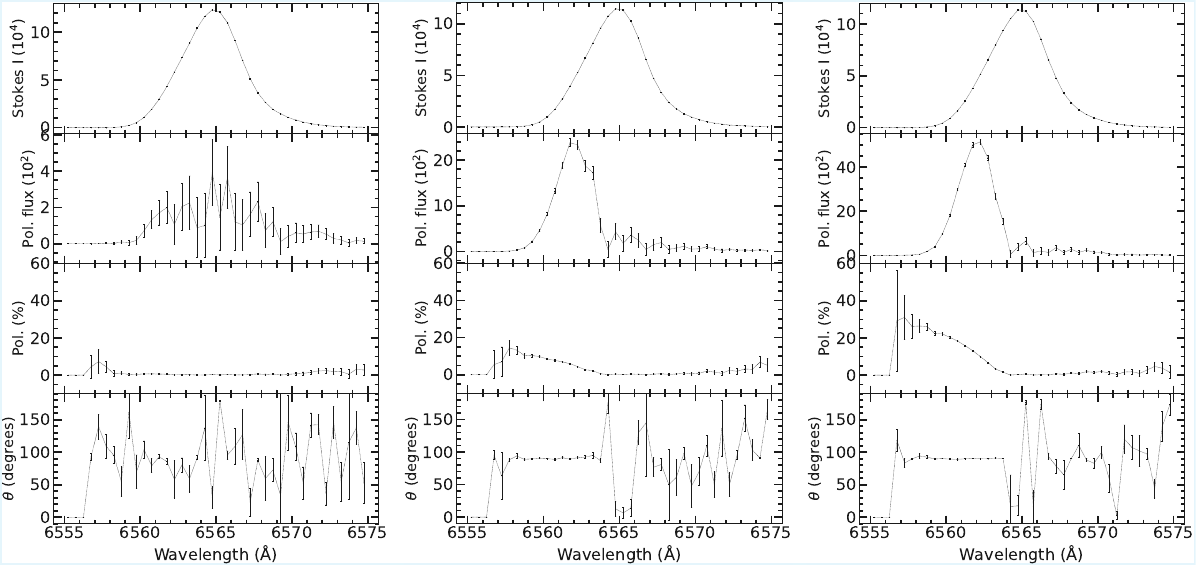} 
  \caption{The figure shows the results of Monte-Carlo simulations of the Raman scattering of Lyman $\beta$ photons (emitted near the WD and scattered in the neutral hydrogen wind of the red giant) for a typical symbiotic system with stellar parameters similar to Y Gem (See text for details), considering an edge-on system. The polarization profiles are simulated for the phases that correspond to the epochs of spectro-polarimetric observation, viz. March 2024 (left), March 2025 (middle), and December 2025 (right). All simulations have been conducted with $10^8$ photons.}
  \label{Fig-Symbiotic_MonteCarlo}
\end{figure*}

Figure \ref{Fig-Symbiotic_MonteCarlo} shows the simulated polarization profiles of Y Gem system when viewed edge-on (inclination 90$\degree$), corresponding to the three epochs of spectro-polarimetric observations. The full details of how the simulation was constructed are discussed in Appendix~\ref {subsec-MonteCarlo}. The simulations reveal that almost zero polarization is obtained when the system is viewed along the binary axis, while maximum polarization is obtained at quadrature. The non-detection of polarization signature in March 2024 observations indicates that the binary axis of the Y Gem system was oriented along the observer's line of sight. Similar Monte Carlo simulations to simulate the Raman scattering of O VI $\lambda \lambda$ $1032, 1038 \text{\AA}$ resonance doublet photons in the neutral hydrogen wind of the red giant had also led to the same conclusion that the integrated polarization across the Raman scattered $\lambda \lambda$ $6830, 7088 \text{\AA}$ features attains a minima when the system is viewed along the binary axis and maxima when the system is viewed at quadrature \citep{schmid1992montecarlo, schmid1995monte, schmid1996simulations, harries1997raman}.  Thus, in the epoch of first spectro-polarization observation (March 2024), the binary axis must be close to the line of sight, and this, when used in conjunction with the previously discussed epoch of eclipse (December 2020), constrains the possible values of eccentricity of the system. We have, therefore, used the eccentricity value of 0.4 and angle of rotation 130$\degree$ to construct the orbit, though it must be mentioned that other close-by values of the corresponding parameters could also be possible. As stated above, we do not have a sufficient number of observational data points to optimize the true eccentricity value of the system orbit. We can only estimate the probable upper limit of the system's eccentricity using Eggleton's approximation \citep{eggleton1983approximations}, which depends on the mass ratio and the binary separation of the system (see Appendix~\ref{subsec-MonteCarlo}). With these values, the phase 0.0 corresponds to the WD eclipsed by the red giant, whereas the spectro-polarimetric observations on March 2024 correspond to the WD in front of the red giant, with the binary axis making an angle of 7.3 degrees with the observer's line of sight. This would mean that for spectro-polarimetric observations on March 2025 and December 2025, the binary axis would make angles of 27.8 and 42.7 degrees, respectively, with the line of sight of the observer. A significant increase in polarization was predicted as the binary components moved towards quadrature (see Figure~\ref{Fig-Symbiotic_MonteCarlo}), and the same has been observed over the last two epochs (Figure~\ref{Fig-PolProfile}).
\par
As noted above, data for few orbital cycles are needed to firmly establish the orbital period of the binary system. However, for a long-period binary system like Symbiotic stars, the periodic observations (spanning a good fraction of the orbit) may be helpful to put some constraints in order to predict the anticipated orbital phases, when the suitable observations could be made. The orbital model of the scattering geometry (and hence polarization profile) predict significant change in the degree of polarization when both the components of the binary system are along the line of sight of observations compare to when they are at quadrature. Further, if there is a possibility of inferior conjunction /eclipse (in high inclination edge-on systems), it could also be predicted wherein the significant reduction in the strength of emission lines originating from eclipsed components such as H$\alpha$ emission from/in the vicinity of hotter components. In the context of Y Gem, additional follow-up observations of the system every 2-3 months for next few years should provide sufficient phase coverage around the orbit to fully establish an ephemeris of the system. If the adopted model is correct, the spectro-polarimetric data suggest the system is currently moving towards quadrature position, and the model predicts the occurrence of an eclipse event around 2029 (i.e. 8.87 years from December 2020) where significant reduction/disappearance of H$\alpha$ emission could be noticed.

\begin{figure*}
  \centering
  \includegraphics[width=0.8\textwidth]{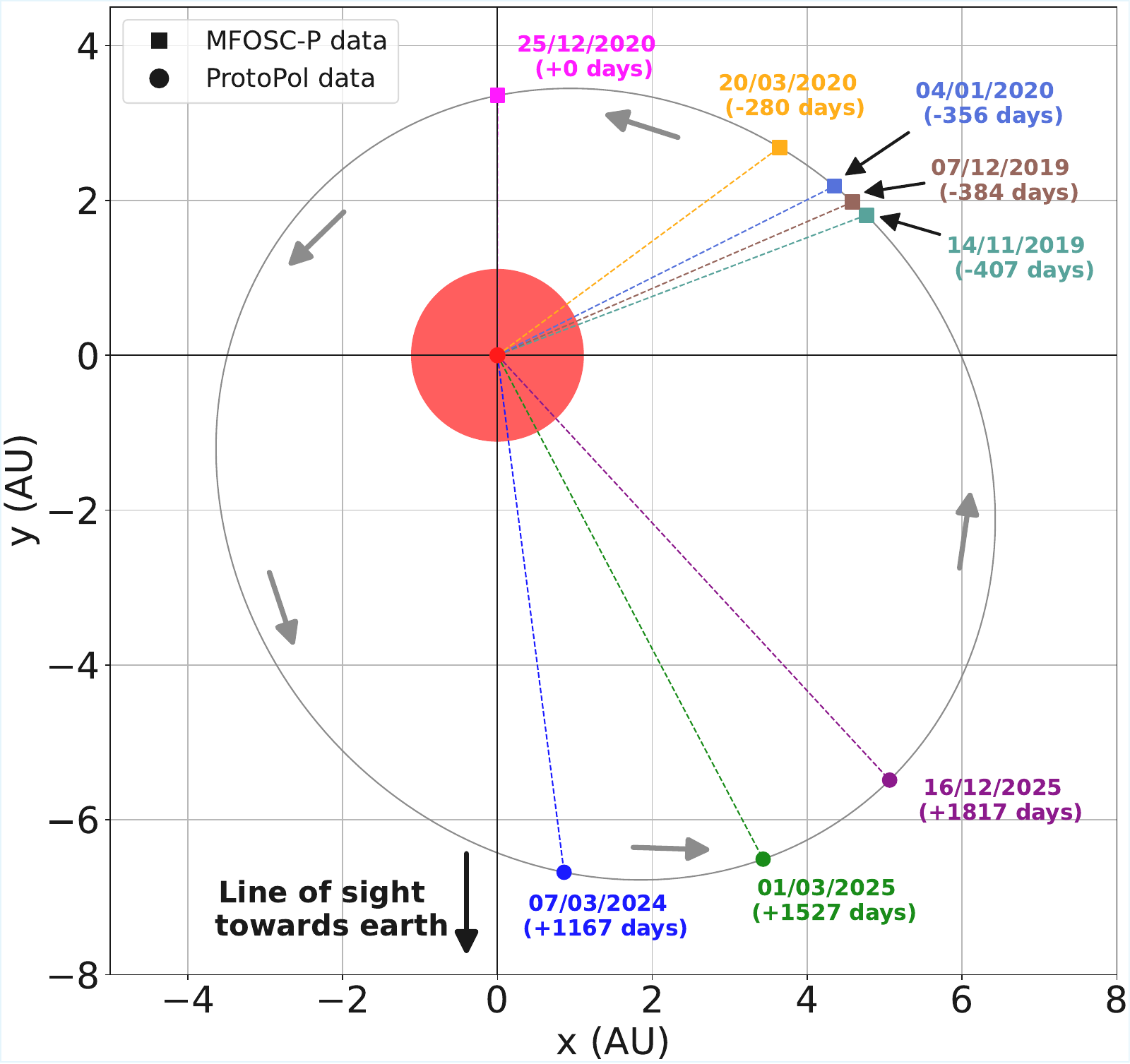} 
  
  \caption{The simple orbital-phase model of Y Gem symbiotic star system (in the reference frame of the red giant) at different epochs of observation as simulated from 2-body dynamics. The red circle shows the red giant of radius 1.116 AU. The magenta square and the blue, green, and purple dots show the position of the WD at epochs 25 December 2020 (H$\alpha$ disappearance), 06 March 2024 (No detected polarization across H$\alpha$), 01 March 2025 (Detected polarization across H$\alpha$), and 16 December 2025 (Increased polarization across H$\alpha$), respectively. The observation epochs of low-resolution spectroscopy are marked by squares, whereas the observation epochs of spectro-polarimetry are marked by circles. The observer is viewing the system from -y-axis, while the dotted lines represent the corresponding binary axis at each epoch. The 2-body simulation is generated in the red giant's reference frame with red giant and WD masses of 1.1 $M_\odot$ and 0.8 $M_\odot$, respectively. The semi-major axis of the system is 5.3 AU, and an eccentricity of 0.4 is assumed for the system. The orbital period of the system was taken as 8.87 years. All system parameters (except eccentricity and angle of rotation) were adopted from a recent work on Y Gem \citep{guerrero2025gem}. See section
  ~\ref{subsec-OrbitalPhase} for discussion.}
  \label{Fig-phase_geometry}
\end{figure*}

\section{Conclusion}
\label{Sec-conclusions}

The spectro-polarimetric observations of H$\alpha$, presented in this work, most likely offer the strongest case for the variation of Raman-scattered Ly$\beta$ features with the orbital phase. The strong polarization seen in the H$\alpha$ at certain phases and its significant deviation from other phases are strong observational evidence to validate the astrophysical models of symbiotic systems. Utilizing the medium-resolution spectro-polarimetric data and the low-resolution spectroscopic data, we have attempted to explain the observed variability in the polarization profile of H$\alpha$ due to the morphological changes induced by the orbital-phase. A simple orbital model of Y Gem is constructed to put constraints on the inclination angle and phase of the symbiotic star system which hints towards the inclination of the system to be close to edge-on. We emphasize that the model presented here is not a unique orbital solution but is rather used as a framework to establish the cause of observed variability in polarization as the orbital phase-induced scattering of Ly$\beta$ photons. The present work also emphasizes the significance and utility of multi-epoch spectro-polarimetric observations in deciphering the orbital solutions and the morphological constructs.
\par

It has long been suggested that such spectro-polarimetric observations could be used to construct the orbit of the binary systems \citep{schmid1992montecarlo, harries1997raman, lee1997profiles}. And, indeed, some such attempts were made \citep{schild1996spectropolarimetry, harries1996spectropolarimetric, schmid1997spectropolarimetry} wherein the strongly polarized and well resolved O VI Raman scattered features were used to construct the orbits. However, such attempts were mostly limited to symbiotic stars only, as these features are only seen in symbiotics. Thus, the detection and variation of the strong polarization in the H$\alpha$ profile, most likely caused by the Raman-scattered Ly$\beta$ photons, opens a promising window for a variety of astrophysical situations, due to the requirements of low ionization conditions, high scattering cross sections, etc., which are in contrast to the input physical conditions needed for O VI photons. It is shown here that the detection of Raman-scattered Ly$\beta$ photons, coupled with modeling efforts, provides a powerful diagnostic tool to probe the physical conditions and scattering geometry associated with the symbiotic star environment, and the same can be applied to other astrophysical situations as well. 
\par 
Raman-scattered Ly$\beta$ features provide a powerful diagnostic of physical conditions in astrophysical sources. Their strengths and profiles depend sensitively on the neutral hydrogen column density, the geometry between the UV source and the scattering region, and the velocity field of the gas, etc. \citep{jung2004centre, lee2012analysis}. It is imperative to compare the Raman scattered features of Ly$\beta$ with those of the O VI doublet (1032 and 1038 $\AA$). The O VI double arises from much higher ionization conditions and probes a markedly different physical regime. As the ionization potential (IP) of O V is around 113.9 eV, the O VI photons are, thus, produced in extremely hot plasma (T $\ge$ 10$^{5}$ K), typically close to compact objects or in fast shocks. The Ly$\beta$ photons, on the other hand, originate from hydrogen recombination or resonance processes in moderately ionized gas. Further, as the scattering cross-section sharply increases near the Ly$\beta$ wavelength, Raman scattering of Ly$\beta$ photons becomes significant in neutral regions with column density (N$_{HI}$) $\ge 10^{20}$ cm$^{-2}$, which is two-orders of magnitude smaller than the typical column densities associated with Raman scattering of O VI photons \citep{yoo2002polarization}. This hints at the fact that Ly$\beta$ photons are scattered in a more extended wind of the red giant, whereas the O VI photons are predominantly scattered in the regions where the scatterer densities are higher. This is also indicated by the presence of broad H$\alpha$ wings in most symbiotic stars, whereas only half of them show Raman scattered O VI  features \citep{van1993atlas, ivison1994atlas}. Thus, Raman scattering of Ly$\beta$ and O VI photons provides complementary constraints on astrophysical media, enabling a more complete reconstruction of ionization structure, kinematics, mass-transfer processes, and morphology of the scattering media in complex interacting systems. Furthermore, broad H$\alpha$ wings are observed in several other astrophysical systems like young planetary nebulae, post-AGB stars, AGNs, etc., where Raman scattering of Ly$\beta$ has been stated as a probable cause of line broadening \citep{arrieta2003broad, chang2015formation}. Therefore, the multi-phase spectro-polarimetric observations of such astrophysical systems may prove to be a powerful diagnostic tool.


\section*{Data Availability}

The data presented in this manuscript may be made available upon reasonable request. Corresponding authors may be contacted for that.


\begin{acknowledgments}
We thank the anonymous reviewers for their review and constructive comments and suggestions
on the paper. The research work at the Physical Research Laboratory (PRL), Ahmedabad, is funded by the Department of Space (DOS), Govt. of India. AM gratefully acknowledges PRL for a Ph.D. research fellowship. PRL operates the Mt. Abu observatory with 1.2 m and 2.5 m telescopes at Mt. Abu. We acknowledge the use of data collected from both the PRL 1.2m and 2.5m telescopes at Mt. Abu Observatory with the MFOSC-P and ProtoPol instruments, respectively.
\end{acknowledgments}

\begin{contribution}

AM was responsible for the observation, data reduction, and analysis with ProtoPol. He developed the code for the Monte-Carlo simulations to simulate the Raman scattering phenomenon in Symbiotic star systems. He led the manuscript preparation. MKS is the PI of ProtoPol and MFOSC-P instrument. He led the project with the initial research concept, supervision of the project, discussion, manuscript editing, etc. VK took the MFOSC-P observations of Y Gem and performed the subsequent data reduction, participated in the discussion, and manuscript preparation.

\appendix

\section{H$\alpha$ spectral deblending}
\label{subsec-SpectralDeblending}

\begin{figure}
    \includegraphics[width=\columnwidth]{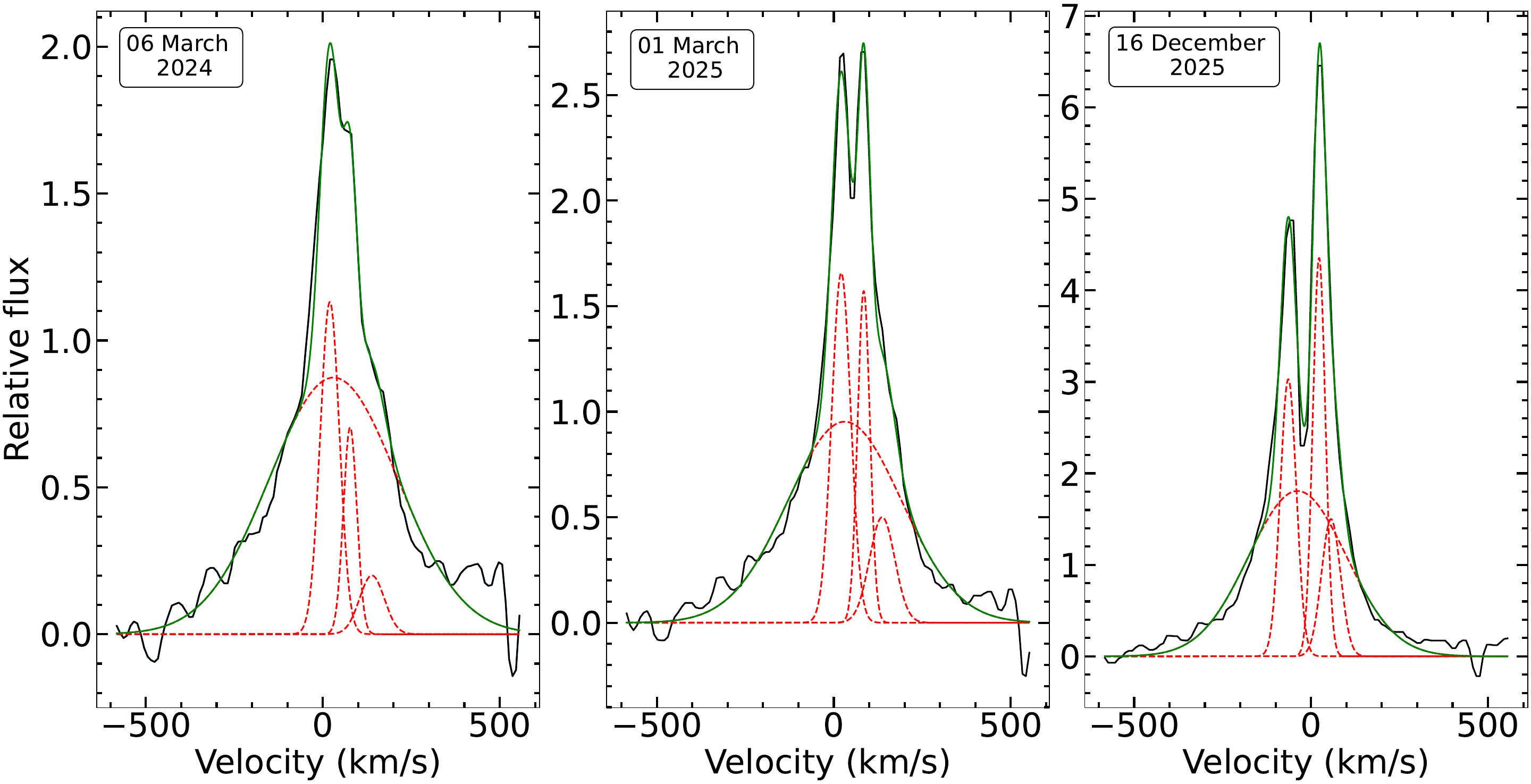}
    \caption{Figures show the decomposed profiles of H$\alpha$ emission of Y Gem as observed in March 2024 \textit{(left)}, March 2025 \textit{(center)}, and December 2025 \textit{(right)} with ProtoPol in velocity space. The individual emission components are shown in red dashed lines, while the combined profiles are in green, overlaid on the observed profiles in black. The spectra are normalized with respect to the local continuum. The H$\alpha$ line profile for each epoch is decomposed into a red and a blue-shifted emission component and a broader component contributing to the flux in the H$\alpha$ wings. A distinct additional emission feature is also noticed in the red part of the profile, causing the asymmetry in the H$\alpha$ profile.}
    \label{Fig-Halpha_deblend}
\end{figure}

\begin{table}
	\centering
	\caption{Parameters derived from the decomposition of H$\alpha$ profile as shown in Figure~\ref{Fig-Halpha_deblend}. The blue-shifted component, the red-shifted component, the underlying broad component, and the emission feature on the redder side are labeled blue, red, broad, and other, respectively.}
	\label{Halpha_deblend_table}
	\begin{tabular}{cccc} 
		\hline
		\textbf{Line name} & \textbf{Relative} & \textbf{Line centre} & \textbf{FWHM}\\
         &  \textbf{Amplitude} & \textbf{(km s$^{-1}$)} & \textbf{(km s$^{-1}$)}\\
		\hline
        
         & \multicolumn{2}{c|}{\textbf{Epoch 1: March 2024}} \\
        Blue & 1.130 $\pm$ 0.071 & 20.000 $\pm$ 4.013 & 63.898 $\pm$ 8.353\\
        Red & 0.704 $\pm$ 0.190 & 77.234 $\pm$ 4.376 & 44.957 $\pm$ 11.944\\
        Broad & 0.873 $\pm$ 0.048 & 30.227 $\pm$ 6.001 & 427.455 $\pm$ 14.464\\
        Other  & 0.200 $\pm$ 0.065 & 138.806 $\pm$ 27.603 & 82.425 $\pm$ 56.160\\
        
         & \multicolumn{2}{c|}{\textbf{Epoch 2: March 2025}} \\
        Blue & 1.656 $\pm$ 0.064 & 21.020 $\pm$ 1.421 & 63.299 $\pm$ 3.999\\
        Red & 1.571 $\pm$ 0.213 & 84.640 $\pm$ 1.377 & 40.862 $\pm$ 4.258\\
        Broad & 0.952 $\pm$ 0.054 & 30.903 $\pm$ 5.600 & 378.615 $\pm$ 12.348\\
        Other  & 0.500 $\pm$ 0.078 & 136.714 $\pm$ 14.236 & 83.971 $\pm$ 26.295\\

        & \multicolumn{2}{c|}{\textbf{Epoch 3: December 2025}} \\
        Blue & 3.097 $\pm$ 0.145 & -64.454 $\pm$ 0.886 & 54.015 $\pm$ 2.975\\
        Red & 5.065 $\pm$ 0.198 & 26.207 $\pm$ 1.480 & 47.771 $\pm$ 2.608\\
        Broad & 1.731 $\pm$ 0.133 & -42.099 $\pm$ 5.909 & 327.983 $\pm$ 13.573\\
        Other  & 0.919 $\pm$ 0.157 & 80.000 $\pm$ 9.144 & 54.274 $\pm$ 17.015\\
        
		\hline
	\end{tabular}
\end{table}

\par
The observed H$\alpha$ line profiles as observed with ProtoPol over the three observation epochs in velocity space are shown in Figure~\ref{Fig-Halpha_deblend} with two narrow red-shifted and blue-shifted components and one broad component underneath. Another distinct additional emission feature is also observed in the red part of the profile in all the epochs. We shall not discuss this additional component in the present context, as its origin could be external to the Y Gem system, such as a physically distinct cloud of hydrogen in velocity space, or another unidentified emission line, etc. On the first and second epochs of observations in March 2024 and 2025 (nearly 1 year apart), the red and blue-shifted peaks have similar FWHMs and position in velocity space, respectively; however, the relative amplitudes of the peaks change considerably, and a transition from an almost single peak profile to an equi-strength double peak profile is clearly noticed. The peak-to-peak separation of the two components also remains constant at approximately 60 km s$^{-1}$. The FWHM of the broad-line component varied between 370 and 430 km s$^{-1}$. However, at the third epoch in December 2025, the entire profile appears to be blue-shifted by about 60 km s$^{-1}$, though admittedly, no Heliocentric correction has been applied to the observed spectra. The peak-to-peak separation of blue and red peaks has increased to around 80 km s$^{-1}$. The amplitudes of these components with respect to the continuum have also increased significantly as compared to the first two epochs, with the red peak gaining a slightly higher value. These values are typical of the symbiotic systems, for example, symbiotic star AG Peg shows similar values \citep{lee2018h, hyung2020formation}.  The exact values of the decomposed lines are given in table \ref{Halpha_deblend_table}.

\section{Monte Carlo modeling of the Raman Scattering of Ly$\beta$ photons}
\label{subsec-MonteCarlo}


A full Stokes Monte-Carlo radiative transfer code has been developed to model the  Raman scattering of Ly$\beta$ $\lambda$1025.7 $\AA$ photons in the neutral hydrogen wind of the red giant star in a symbiotic binary. Few attempts, at simulating the Raman scattering of Ly$\beta$ photons, have been made through various approaches/geometries in the literature, mostly to explain the broad wings of H$\alpha$ seen in various symbiotics e.g., a two-component scattering region with a cylindrical shell coupled to a bipolar finite slab \citep{yoo2002polarization}; a partial spherical shell as the neutral scattering region with the source at the center of the sphere \citep{chang2018broad} etc. However, here we have attempted to simulate the polarization profile caused by such scattering and matching them with the observational signatures. Our model is based on similar works to simulate the polarization profile of Raman scattering of O VI resonance doublet (O VI $\lambda \lambda$ $1032, 1038 \text{\AA}$) photons in the neutral wind of the red giant\citep{harries1997raman, schmid1992montecarlo, schmid1995monte, schmid1996simulations, lee1997profiles}. 
\par 
Here, we have adopted a system geometry that consists of a red giant star of radius $R_G$, whose center coincides with the origin of the global coordinate system. The hot component is located at a separation of $q_{sep}$ from the red giant's center on the x-axis. Ly$\beta$ photon packets are launched in an isotropic way from the hot component and are assigned initial frequencies by applying a Doppler shift corresponding to the intrinsic Lyman $\beta$ line width and by approximating the line profile as a Gaussian distribution. The photons propagate through a spherically symmetric, constant-velocity red-giant wind, where the wind density follows the $n(r) \propto 1/r^2$ law. Using this form of the wind density allows for the computation of the scattering optical depths analytically for this geometry, which is then used to compute the path length to the next scattering event.
\par
For each of the scattering event, the interaction is considered as Rayleigh or Raman in nature, whose corresponding probability is taken to be the ratio of the relevant cross-sections ($\sigma$) e.g. probability for Raman scattering to occur is $P_{Raman} = \sigma_{Raman}/(\sigma_{Rayleigh} + \sigma_{Raman})$. The Raman scattering cross-sections, as given by Kramers–Heisenberg formula, are determined using the equation $\sigma_{Raman} = K \frac{\omega_{32}}{\omega - \omega_{31}} \sigma_T$, where $\sigma_T = 6.6 \times 10^{-25}$ cm$^{2}$ is the Thompson scattering cross-section, $\omega_{32},\: \omega_{31}$ are the angular frequencies associated with hydrogen transition from 3$\to$2 and 3$\to$1 level, and K is a constant \citep{chang2018broad}. The branching ratio ($ r_b =  \sigma_{Raman}/\sigma_{Rayleigh}$) is itself a function of the emitted photon wavelength and is approximated by $r_b(\omega) = 0.1342 - 12.5((\omega - \omega_{\beta})/\omega_{\beta})$ where $\omega_{\beta}$ is the angular frequency corresponding to Ly$\beta$ wavelength \citep{yoo2002polarization}. Using the branching ratio and the Raman scattering cross-section values, the Rayleigh scattering cross-section for the particular wavelength of the photon is calculated. The Rayleigh events are treated as elastic scattering events but incorporate Doppler shift effects due to the local bulk velocity of the wind. Raman events convert the photon to the optical wavelength via the equation $\nu_{Raman} = \nu_{in} - \nu_{Ly\alpha}$, where $\nu_{Raman}$, $\nu_{in}$, and $\nu_{Ly\alpha}$ are the Raman, incoming and Lyman $\alpha$ frequencies, respectively. While Rayleigh scattering events change the direction of propagation of the photons in the scattering media, as soon as a Raman scattering event is encountered, the photon is scattered directly towards the observer, assuming the media to be transparent to optical photons \citep{harries1997raman}. 
\par
Each scattering event updates the Stokes vector (I, Q, U) associated with the photon (initially assumed to be unpolarized during photon creation) using the full dipole-scattering Mueller matrix formulation \citep{harries1997raman, kokubo2024rayleigh}. An orthogonal polarization basis is constructed to the direction of propagation of the photon and referenced to the global +Z axis in the observer’s sky plane, enabling consistent tracking of polarization through arbitrary scattering geometries. The photon is terminated within the simulation when a Raman scattering phenomenon occurs or when the photon escapes from the system boundary (taken as the radius of 1000$R_G$). For photons that have successfully Raman scattered, their emergent wavelength and final Stokes parameters are recorded. These data are accumulated into spectral bins to produce synthetic intensity, polarized flux, degree of polarization, and polarization angle spectra. The above process is simulated for various inclination angles and phases of the binary system, w.r.t the observer.
\par
To simulate the polarization profile of the Y Gem system using the simulation model discussed above, the following parameters were used. The masses of the red giant and WD are taken as 1.1 $M_\odot$ and 0.8 $M_\odot$, respectively,  the radius of the red giant as  240 $R_\odot$, and the semi-major axis of the system as 5.3 AU. These values are suggested in a recent work on Y Gem \citep{guerrero2025gem}. The system's eccentricity was estimated by Eggleton's approximation \citep{eggleton1983approximations}, which allows us to restrict the maximum value of the eccentricity that a binary stellar system can have. For the system configuration assumed above for Y Gem, this maximum eccentricity turns out to be 0.48. The eccentricity is later visually optimized to simulate the polarization profiles similar to the observed ones as per orbital phases (see section~\ref{subsec-OrbitalPhase}) and is, thus, determined to be 0.4. The mass-loss rate of 10$^{-7}$  M$_\odot$yr$^{-1}$ and wind velocity of 15 kms$^{-1}$ are taken as the typical values of S-type symbiotic systems like Y Gem. The full width at half maximum (FWHM) of the Ly$\beta$ line in velocity space is taken as 20 kms$^{-1}$. We must clarify that the parameters adopted here are only the indicative values for a symbiotic system, and we did not attempt to fully simulate the physical parameter space of Y Gem. Our only attempt here is to demonstrate that, given a suitable and possible set of parameters, it is possible to explain the observed polarization profile of H$\alpha$ as the one caused by Raman scattering of Ly$\beta$ photons. 
\par 
The polarization profiles are then generated from the simulation model for the three different epochs of spectro-polarimetric observations, viz. March 2024, March 2025, and December 2025. The optimized orbital phase model provides the binary separation (6.73, 7.36, and 7.47 AU, respectively) and angle subtended by the binary axis w.r.t the line of sight of the observer are (7.3$\degree$, 27.8$\degree$, and 42.7$\degree$). The simulations have been performed to check the polarization profiles at various inclination angles of the orbital plane w.r.t the observer as well as various orbital phases. The face-on orbit showed the least variation in the polarization profiles w.r.t orbital phase, whereas the edge-on orbit showed the maximum changes in the polarization profiles. This was consistent with the similar results from the simulations of Raman scattered O VI lines \citep{schmid1992montecarlo, harries1997raman, lee1997profiles}. The simulations were performed iteratively to produce polarization profiles similar to the observed ones, by fixing the observer's line of sight and varying the inclination angle and orbital phase of the system. As the simulations predict, during the first epoch (March 2024), when the WD is almost in front of the red giant along the line of sight, the polarization is almost negligible. However, as the system moves towards quadrature, the polarization increases, firstly in the March 2025 epoch and even more so in the December 2025 epoch. The angle of polarization provides another critical piece of evidence in the form of the 90$\degree$ polarization-flip from blue-to-red wings of the H$\alpha$ profile, which are clearly observed in March 2025 and December 2025 (more prominent) and have successfully been simulated by the model. The simulations confirm the presence of multiple scattering regions and mirror symmetry of the scattering media with respect to the orbital plane of the symbiotic system as the cause of the observed polarization angle flips.

\section{Additional Observations}
\label{subsec-SupplementaryData}

Since the submission of the original manuscript in early March 2026, Y Gem has been observed for a further two epochs, namely, 20 March 2026 and 11 April 2026, though in rather poorer observing conditions (partially cloudy and hazy skies). Therefore, the Stokes I parameter (total flux) in H$\alpha$ is nearly an order of magnitude less than the data presented in the original manuscript. Thus the continuum polarization data on both sides of the H$\alpha$ feature is not well constrained due to a lack of signal, as can be seen in Figure~\ref{Fig-PolProfile_supplementary}, and have higher errors. Nevertheless, the polarization at the H$\alpha$ emission is still noticeable, though with lesser amplitude ($\sim1.2\%$) as compared to the observations taken in December 2025. The deblending of the emission feature is also performed as in the previous epochs and are shown in Figure~\ref{Fig-Halpha_deblend_supplementary}. The parameters of the deblended components are stated in Table~\ref{Table-Halpha_deblend_table_supplementary}. The FWHM of the deblended features are similar to what was derived for the previous epochs.

\begin{table}
	\centering
	\caption{Parameters derived from the decomposition of H$\alpha$ profile as shown in Figure~\ref{Fig-Halpha_deblend_supplementary}. The blue-shifted component, the red-shifted component, the underlying broad component, and the emission feature on the redder side are labeled blue, red, broad, and other, respectively.}
	\label{Table-Halpha_deblend_table_supplementary}
	\begin{tabular}{cccc} 
		\hline
		\textbf{Line name} & \textbf{Relative} & \textbf{Line centre} & \textbf{FWHM}\\
         &  \textbf{Amplitude} & \textbf{(km s$^{-1}$)} & \textbf{(km s$^{-1}$)}\\
		\hline
        
         & \multicolumn{2}{c|}{\textbf{Epoch 4: March 2026}} \\
        Blue & 2.679 $\pm$ 0.134 & -80.000 $\pm$ 1.075 & 57.107 $\pm$ 3.479\\
        Red & 3.665 $\pm$ 0.199 & 5.163 $\pm$ 2.267 & 43.352 $\pm$ 3.960\\
        Broad & 1.17 $\pm$ 0.115 & -50.000 $\pm$ 7.044 & 353.250 $\pm$ 20.938\\
        Other  & 0.854 $\pm$ 0.211 & 50.000 $\pm$ 9.831 & 43.538 $\pm$ 16.521\\
        
         & \multicolumn{2}{c|}{\textbf{Epoch 5: April 2026}}\\
        Blue & 3.643 $\pm$ 0.130 & -40.000 $\pm$ 0.871 & 61.472 $\pm$ 2.794\\
        Red & 6.135 $\pm$ 1.039 & 50.000 $\pm$ 3.519 & 52.843 $\pm$ 3.943\\
        Broad & 1.663 $\pm$ 0.116 & -13.850 $\pm$ 5.629 & 370.198 $\pm$ 14.534\\
        Other  & 1.000 $\pm$ 0.709 & 94.749 $\pm$ 32.250 & 63.415 $\pm$ 36.498\\
        
		\hline
	\end{tabular}
\end{table}

\begin{figure}
    \centering
    \includegraphics[width=0.9\textwidth]{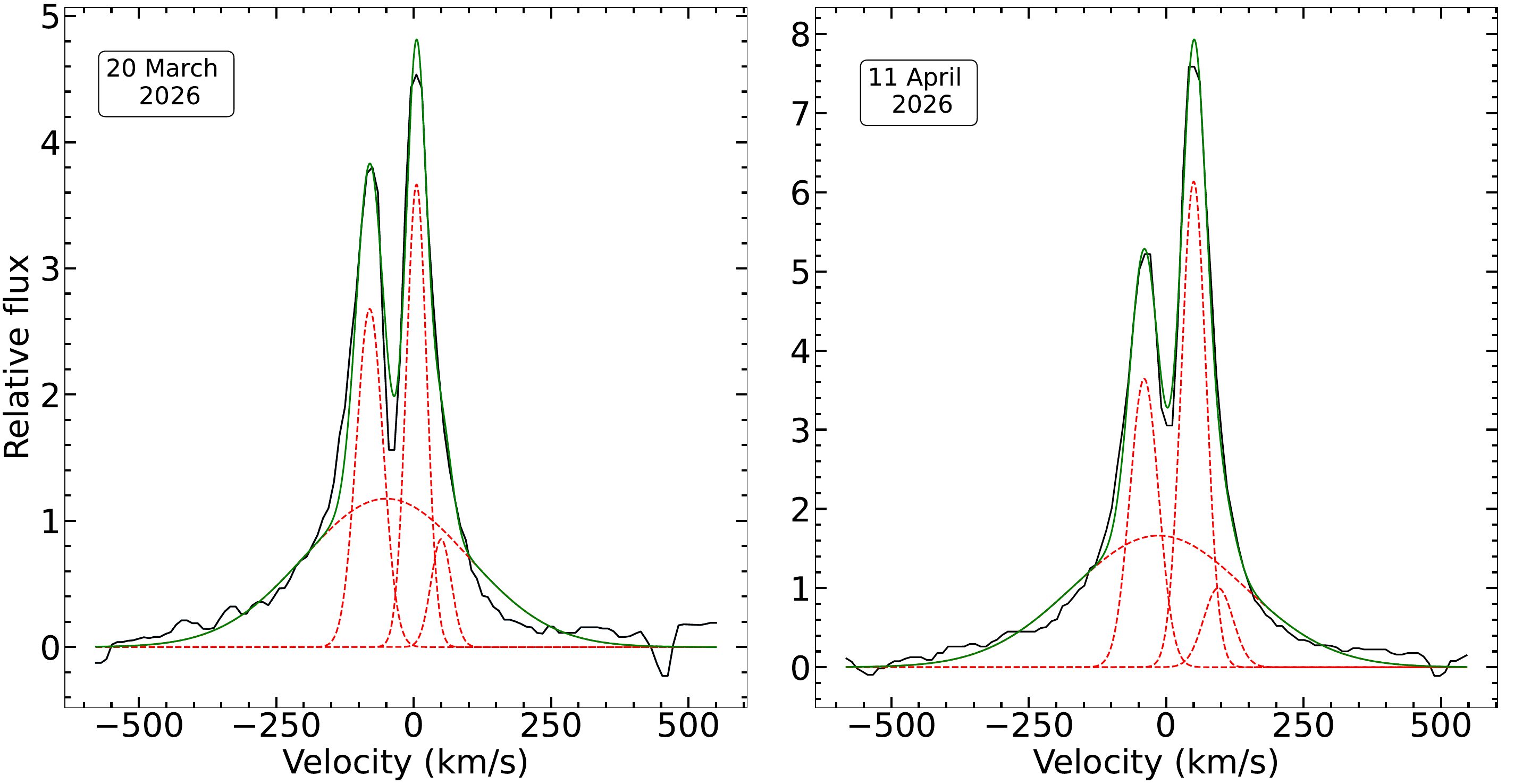}
    \caption{Figures show the decomposed profiles of H$\alpha$ emission of Y Gem as observed in March 2026 \textit{(left)} and April 2026 \textit{(right)} with ProtoPol in velocity space. The individual emission components are shown in red dashed lines, while the combined profiles are in green, overlaid on the observed profiles in black. The spectra are normalized with respect to the local continuum. The H$\alpha$ line profile for each epoch is decomposed into a red and a blue-shifted emission component and a broader component contributing to the flux in the H$\alpha$ wings. A distinct additional emission feature is also noticed in the red part of the profile, causing the asymmetry in the H$\alpha$ profile.}
    \label{Fig-Halpha_deblend_supplementary}
\end{figure}


\begin{figure*}
  \centering
  \includegraphics[width=0.9\textwidth]{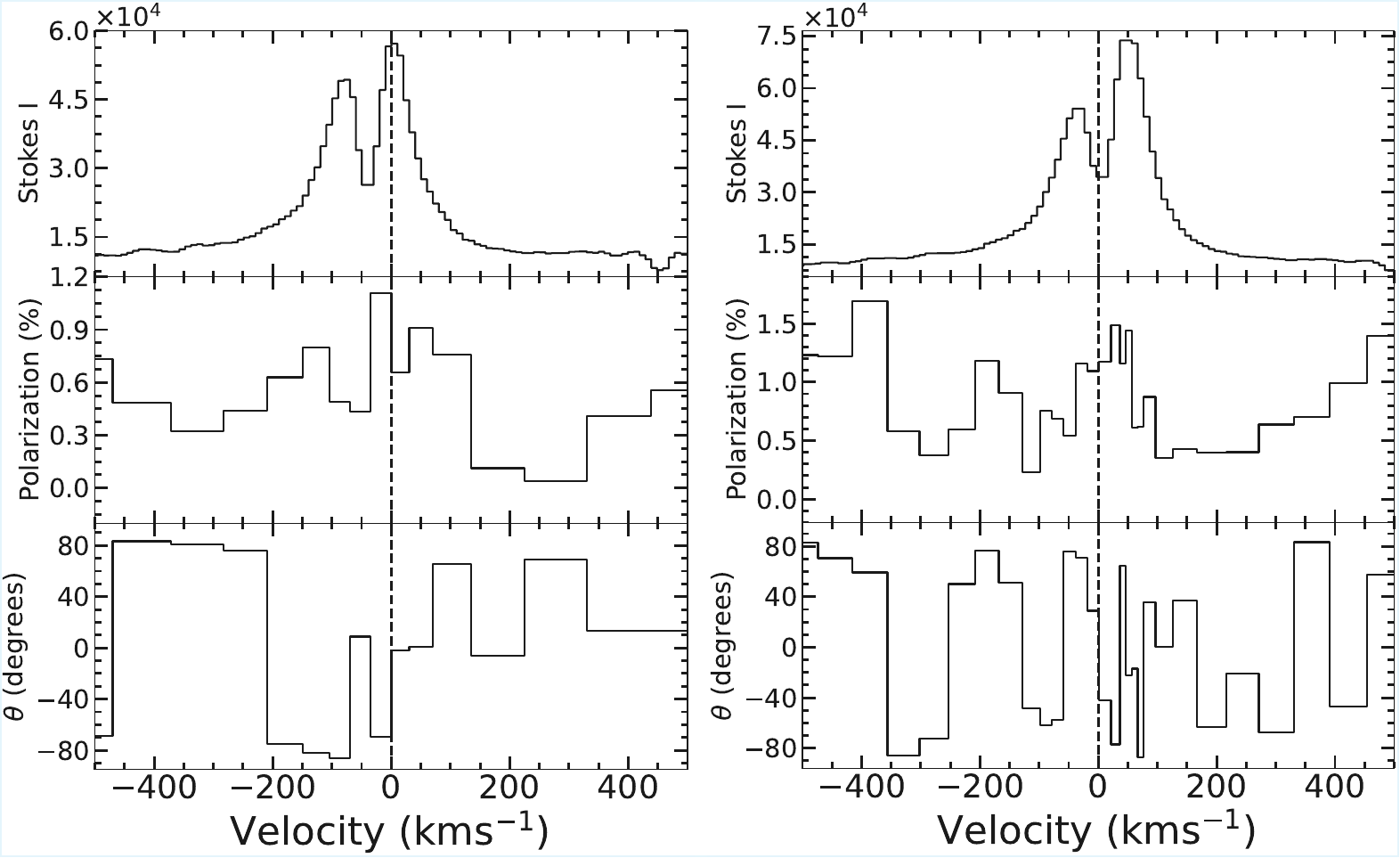} 
  \caption{Polarization profiles for H$\alpha$ as recorded in the spectra of symbiotic star Y Gem for March 2026 (left) and April 2026 (right). The data for the two observation dates have been binned to achieve a polarization accuracy of $0.2\%$ and $0.30\%$, respectively, per binning element.}
  \label{Fig-PolProfile_supplementary}
\end{figure*}



\end{contribution}

\bibliography{sample701}{}

@article{hyung2020formation,
  title={The Formation of the Double Gaussian Line Profiles of the Symbiotic Star AG Peg},
  author={Hyung, Siek and Lee, Seong-Jae},
  journal={arXiv preprint arXiv:2002.06500},
  year={2020}
}

@article{ikeda2004polarized,
  title={Polarized H$\alpha$ Wings in the Symbiotic Stars AG Draconis and Z Andromedae},
  author={Ikeda, Yuji and Akitaya, Hiroshi and Matsuda, Kentaro and Homma, Ken’ichi and Seki, Munezo and Kawabata, Koji S and Hirata, Ryuko and Okazaki, Akira},
  journal={The Astrophysical Journal},
  volume={604},
  number={1},
  pages={357},
  year={2004},
  publisher={IOP Publishing}
}

@article{chang2018broad,
  title={Broad Wings around H$\alpha$ and H$\beta$ in the Two S-type Symbiotic Stars Z Andromedae and AG Draconis},
  author={Chang, Seok-Jun and Lee, Hee-Won and Lee, Ho-Gyu and Hwang, Narae and Ahn, Sang-Hyeon and Park, Byeong-Gon},
  journal={The Astrophysical Journal},
  volume={866},
  number={2},
  pages={129},
  year={2018},
  publisher={IOP Publishing}
}

@article{yoo2002polarization,
  title={Polarization of the broad H$\alpha$ wing in symbiotic stars},
  author={Yoo, Jerry Jaiyul and Bak, Jih-Yong and Lee, Hee-Won},
  journal={Monthly Notices of the Royal Astronomical Society},
  volume={336},
  number={2},
  pages={467--476},
  year={2002},
  publisher={Blackwell Science Ltd}
}

@article{robinson1994nature,
  title={On the nature of the emission-line profiles of symbiotic stars--I. Accretion discs},
  author={Robinson, K and Bode, MF and Skopal, A and Ivison, RJ and Meaburn, J},
  journal={Monthly Notices of the Royal Astronomical Society},
  volume={269},
  number={1},
  pages={1--12},
  year={1994},
  publisher={Oxford University Press Oxford, UK}
}

@article{lee2018h,
  title={H $\alpha$ and H $\beta$ Raman scattering line profiles of the symbiotic star AG Pegasi},
  author={Lee, Seong-Jae and Hyung, Siek},
  journal={Monthly Notices of the Royal Astronomical Society},
  volume={475},
  number={4},
  pages={5558--5569},
  year={2018},
  publisher={Oxford University Press}
}

@article{guerrero2025gem,
  title={Y Gem, a symbiotic star outshone by its asymptotic giant branch primary component},
  author={Guerrero, Mart{\'\i}n A and Vasquez-Torres, DA and Rodr{\'\i}guez-Gonz{\'a}lez, JB and Toal{\'a}, Jes{\'u}s A and Ortiz, R},
  journal={Astronomy \& Astrophysics},
  volume={693},
  pages={A203},
  year={2025},
  publisher={EDP Sciences}
}

@article{yu2022gem,
  title={Y Gem: A White Dwarf Symbiotic Star?},
  author={Yu, Zhuo-li and Xu, Xiao-jie and Shao, Yong and Wang, Q Daniel and Li, Xiang-Dong},
  journal={The Astrophysical Journal},
  volume={932},
  number={2},
  pages={132},
  year={2022},
  publisher={IOP Publishing}
}

@article{sahai2018binarity,
  title={Binarity and Accretion in AGB Stars: HST/STIS Observations of UV Flickering in Y Gem},
  author={Sahai, R and Contreras, C S{\'a}nchez and Mangan, AS and Sanz-Forcada, J and Muthumariappan, C and Claussen, MJ},
  journal={The Astrophysical Journal},
  volume={860},
  number={2},
  pages={105},
  year={2018},
  publisher={IOP Publishing}
}

@article{sahai2011strong,
  title={Strong variable ultraviolet emission from y gem: accretion activity in an asymptotic giant branch star with a binary companion?},
  author={Sahai, Raghvendra and Neill, James D and de Paz, Armando Gil and Contreras, Carmen S{\'a}nchez},
  journal={The Astrophysical Journal Letters},
  volume={740},
  number={2},
  pages={L39},
  year={2011},
  publisher={IOP Publishing}
}

@article{sahai2008binarity,
  title={Binarity in cool asymptotic giant branch stars: a GALEX search for ultraviolet excesses},
  author={Sahai, R and Findeisen, K and De Paz, A Gil and Contreras, C S{\'a}nchez},
  journal={The Astrophysical Journal},
  volume={689},
  number={2},
  pages={1274},
  year={2008},
  publisher={IOP Publishing}
}

@article{cardelli1989relationship,
  title={The relationship between infrared, optical, and ultraviolet extinction},
  author={Cardelli, Jason A and Clayton, Geoffrey C and Mathis, John S},
  journal={Astrophysical Journal, Part 1 (ISSN 0004-637X), vol. 345, Oct. 1, 1989, p. 245-256.},
  volume={345},
  pages={245--256},
  year={1989}
}

@article{taylor1984radio,
  title={Radio emission from symbiotic stars-A binary model},
  author={Taylor, AR and Seaquist, ER},
  journal={Astrophysical Journal, Part 1 (ISSN 0004-637X), vol. 286, Nov. 1, 1984, p. 263-268. Research supported by the Natural Sciences and Engineering Research Council of Canada.},
  volume={286},
  pages={263--268},
  year={1984}
}

@article{lee2000raman,
  title={Raman-Scattering Wings of H$\alpha$ in SymbioticStars},
  author={Lee, Hee-Won},
  journal={The Astrophysical Journal},
  volume={541},
  number={1},
  pages={L25},
  year={2000},
  publisher={IOP Publishing}
}

@article{nussbaumer1989raman,
  title={Raman scattering as a diagnostic possibility in astrophysics},
  author={Nussbaumer, H and Schmid, HM and Vogel, M},
  journal={Astronomy and Astrophysics (ISSN 0004-6361), vol. 211, no. 2, March 1989, p. L27-L30. Research supported by SNSF.},
  volume={211},
  pages={L27--L30},
  year={1989}
}

@article{schmid1989identification,
  title={Identification of the emission bands at 6830, 7088 A},
  author={Schmid, HM},
  journal={Astronomy and Astrophysics (ISSN 0004-6361), vol. 211, no. 2, March 1989, p. L31-L34. Research supported by SNSF.},
  volume={211},
  pages={L31--L34},
  year={1989}
}

@article{schmid1994raman,
  title={Raman scattered emission lines in symbiotic stars: a spectropolarimetric survey},
  author={Schmid, HM and Schild, H},
  journal={Astronomy and Astrophysics (ISSN 0004-6361), vol. 281, no. 1, p. 145-160},
  volume={281},
  pages={145--160},
  year={1994}
}

@article{harries1996raman,
  title={Raman scattering in symbiotic stars. I. Spectropolarimetric observations},
  author={Harries, TJ and Howarth, ID},
  journal={Astronomy and Astrophysics Supplement Series},
  volume={119},
  number={1},
  pages={61--90},
  year={1996},
  publisher={EDP Sciences}
}

@article{sekeravs2012electron,
  title={Electron optical depths and temperatures of symbiotic nebulae from Thomson scattering},
  author={Seker{\'a}{\v{s}}, M and Skopal, A},
  journal={Monthly Notices of the Royal Astronomical Society},
  volume={427},
  number={2},
  pages={979--987},
  year={2012},
  publisher={Blackwell Science Ltd Oxford, UK}
}

@article{harries1996accretion,
  title={Accretion-disk reflection of the bipolar wind of BI Crucis.},
  author={Harries, TJ},
  journal={Astronomy and Astrophysics, v. 315, p. 499-509},
  volume={315},
  pages={499--509},
  year={1996}
}

@article{kim2007Thompson,
    title = {A Monte Carlo study of polarization structures in the Thomson-scattered line radiation},
    author = {Kim, Hyo Jeong and Lee, Hee-Won and Kang, Suna},
    journal = {Monthly Notices of the Royal Astronomical Society},
    volume = {374},
    number = {1},
    pages = {187-195},
    year = {2006},
    publisher={Blackwell Science Ltd Oxford, UK},
}

@article{rajpurohit2020first,
  title={First results from MFOSC-P: low-resolution optical spectroscopy of a sample of M dwarfs within 100 parsecs},
  author={Rajpurohit, AS and Kumar, Vipin and Srivastava, Mudit K and Allard, F and Homeier, D and Dixit, Vaibhav and Patel, Ankita},
  journal={Monthly Notices of the Royal Astronomical Society},
  volume={492},
  number={4},
  pages={5844--5852},
  year={2020},
  publisher={Oxford University Press}
}

@article{srivastava2021design,
  title={Design and development of Mt. Abu faint object spectrograph and camera--Pathfinder (MFOSC-P) for PRL 1.2 m Mt. Abu Telescope},
  author={Srivastava, Mudit K and Kumar, Vipin and Dixit, Vaibhav and Patel, Ankita and Jangra, Mohanlal and Rajpurohit, AS and Mathur, SN},
  journal={Experimental Astronomy},
  volume={51},
  number={2},
  pages={345--382},
  year={2021},
  publisher={Springer}
}

@inproceedings{kumar2022designs,
  title={Designs of Mt. Abu faint object spectrograph and camera-echelle polarimeter (M-FOSC-EP) and its prototype: spectro-polarimeters for PRL 1.2 m and 2.5 m Mt. Abu Telescopes, India},
  author={Kumar, Vipin and Srivastava, Mudit K and Dixit, Vaibhav and Mistry, Bhavesh and Lad, Kevikumar and Patel, Ankita and Rajpurohit, Arvind S},
  booktitle={Ground-based and Airborne Instrumentation for Astronomy IX},
  volume={12184},
  pages={1696--1714},
  year={2022},
  organization={SPIE}
}

@article{kumar2022optical,
  title={Optical and near-infrared spectroscopy of Nova V2891 Cygni: evidence for shock-induced dust formation},
  author={Kumar, Vipin and Srivastava, Mudit K and Banerjee, Dipankar PK and Woodward, CE and Munari, Ulisse and Evans, Aneurin and Joshi, Vishal and Dallaporta, Sergio and Page, Kim L},
  journal={Monthly Notices of the Royal Astronomical Society},
  volume={510},
  number={3},
  pages={4265--4283},
  year={2022},
  publisher={Oxford University Press}
}

@inproceedings{srivastava2024development,
  title={Development of ProtoPol: a medium resolution echelle spectro-polarimeter for PRL 1.2 m and 2.5 m telescopes, Mt Abu, India},
  author={Srivastava, Mudit K and Maiti, Arijit and Kumar, Vipin and Mistry, Bhaveshkumar and Patel, Ankita and Dixit, Vaibhav and Lad, Kevikumar},
  booktitle={Ground-based and Airborne Instrumentation for Astronomy X},
  volume={13096},
  pages={577--592},
  year={2024},
  organization={SPIE}
}

@incollection{kenyon1986symbiotic,
  title={Symbiotic stars},
  author={Kenyon, Scott J},
  booktitle={Interacting Binaries},
  pages={179--203},
  year={1986},
  publisher={Springer}
}

@article{munari2019symbiotic,
  title={The symbiotic stars},
  author={Munari, Ulisse},
  journal={The Impact of Binary Stars on Stellar Evolution},
  volume={54},
  pages={77},
  year={2019},
  publisher={Cambridge University Press}
}

@article{skopal2001photometric,
  title={A photometric and spectroscopic study of the eclipsing symbiotic binary AX Persei},
  author={Skopal, A and Teodorani, M and Errico, L and Vittone, AA and Ikeda, Y and Tamura, S},
  journal={Astronomy \& Astrophysics},
  volume={367},
  number={1},
  pages={199--210},
  year={2001},
  publisher={EDP Sciences}
}

@article{schmid1992montecarlo,
  title={Montecarlo simulations of Raman scattered OVI emission lines in symbiotic stars},
  author={Schmid, HM},
  journal={Astronomy and Astrophysics, Vol. 254, NO. FEB (I), P. 224, 1992},
  volume={254},
  pages={224},
  year={1992}
}

@article{schmid1995monte,
  title={Monte Carlo simulations of the Rayleigh scattering effects in symbiotic stars},
  author={Schmid, HM},
  journal={Monthly Notices of the Royal Astronomical Society},
  volume={275},
  number={2},
  pages={227--243},
  year={1995},
  publisher={The Royal Astronomical Society}
}

@article{schmid1996simulations,
  title={Simulations of the Raman-scattered O vI emission lines in symbiotic stars},
  author={Schmid, HM},
  journal={Monthly Notices of the Royal Astronomical Society},
  volume={282},
  number={2},
  pages={511--529},
  year={1996},
  publisher={Blackwell Science Ltd Oxford, UK}
}

@article{harries1997raman,
  title={Raman scattering in symbiotic stars. II. Numerical models},
  author={Harries, TJ and Howarth, ID},
  journal={Astronomy and Astrophysics Supplement Series},
  volume={121},
  number={1},
  pages={15--44},
  year={1997},
  publisher={EDP Sciences}
}

@article{lee1997profiles,
  title={On the profiles and the polarization of Raman-scattered emission lines in symbiotic stars—II. Numerical simulations},
  author={Lee, KW and Lee, Hee-Won},
  journal={Monthly Notices of the Royal Astronomical Society},
  volume={292},
  number={3},
  pages={573--590},
  year={1997},
  publisher={Blackwell Science Ltd Oxford, UK}
}

@article{kokubo2024rayleigh,
  title={Rayleigh and Raman scattering cross-sections and phase matrices of the ground-state hydrogen atom, and their astrophysical implications},
  author={Kokubo, Mitsuru},
  journal={Monthly Notices of the Royal Astronomical Society},
  volume={529},
  number={3},
  pages={2131--2149},
  year={2024},
  publisher={Oxford University Press}
}

@article{contini1997evolving,
  title={The Evolving Structure of AG Pegasi, Emerging from the Interpretation of the Emission Spectra at Different Phases},
  author={Contini, Marcella},
  journal={The Astrophysical Journal},
  volume={483},
  number={2},
  pages={887},
  year={1997},
  publisher={IOP Publishing}
}

@article{lee2012analysis,
  title={An analysis of the symbiotic star Z And line profile},
  author={Lee, Seong-Jae and Hyung, Siek and Lee, Kangwhan},
  journal={Journal of the Korean earth science society},
  volume={33},
  number={7},
  pages={608--617},
  year={2012},
  publisher={The Korean Earth Science Society}
}

@article{mennickent2008remarkable,
  title={The remarkable properties of the symbiotic star AE Circinus},
  author={Mennickent, Ronald and Greiner, J and Arenas, J and Tovmassian, G and Mason, E and Tappert, C and Papadaki, C},
  journal={Monthly Notices of the Royal Astronomical Society},
  volume={383},
  number={3},
  pages={845--856},
  year={2008},
  publisher={Blackwell Publishing Ltd Oxford, UK}
}

@article{sonith2023tcp,
  title={TCP J18224935-2408280: a symbiotic star identified during outburst},
  author={Sonith, LS and Kamath, US},
  journal={Monthly Notices of the Royal Astronomical Society},
  volume={526},
  number={4},
  pages={6381--6390},
  year={2023},
  publisher={Oxford University Press}
}

@article{anderson1980observational,
  title={Observational studies of the symbiotic stars. I-H-alpha profile variations in CH Cygni},
  author={Anderson, Ch M and Oliversen, NA and Nordsieck, KH},
  journal={Astrophysical Journal, Part 1, vol. 242, Nov. 15, 1980, p. 188-194.},
  volume={242},
  pages={188--194},
  year={1980}
}

@article{iijima1988high,
  title={High-dispersion spectroscopy of the symbiotic star AX Per: A binary model with rapid mass accretion},
  author={Iijima, T},
  journal={Astrophysics and space science},
  volume={150},
  number={2},
  pages={235--250},
  year={1988},
  publisher={Springer}
}

@article{schmid1997spectropolarimetry,
  title={Spectropolarimetry of symbiotic stars: AG Draconis.},
  author={Schmid, HM and Schild, H},
  journal={Astronomy and Astrophysics, v. 321, p. 791-802},
  volume={321},
  pages={791--802},
  year={1997}
}

@article{schild1996spectropolarimetry,
  title={Spectropolarimetry of symbiotic stars. On the binary orbit and the geometric structure of V1016 Cygni.},
  author={Schild, H and Schmid, HM},
  journal={Astronomy and Astrophysics, v. 310, p. 211-220},
  volume={310},
  pages={211--220},
  year={1996}
}

@article{harries1996spectropolarimetric,
  title={Spectropolarimetric orbits of symbiotic stars: SY Muscae.},
  author={Harries, TJ and Howarth, ID},
  journal={Astronomy and Astrophysics, v. 310, p. 235-238},
  volume={310},
  pages={235--238},
  year={1996}
}

@article{brown1978polarisation,
  title={Polarisation by Thomson scattering in optically thin stellar envelopes. II-Binary and multiple star envelopes and the determination of binary inclinations},
  author={Brown, JC and McLean, IS and Emslie, AG},
  journal={Astronomy and Astrophysics, vol. 68, no. 3, Aug. 1978, p. 415-427. Research supported by the Science Research Council.},
  volume={68},
  pages={415--427},
  year={1978}
}

@article{rudy1978polarimetric,
  title={A polarimetric determination of binary inclinations-Results for five systems},
  author={Rudy, Richard J and Kemp, James C},
  journal={Astrophysical Journal, Part 1, vol. 221, Apr. 1, 1978, p. 200-210.},
  volume={221},
  pages={200--210},
  year={1978}
}

@article{st1993polarization,
  title={Polarization eclipse model of the Wolf-Rayet binary V444 Cygni with constraints on the stellar radii and an estimate of the Wolf-Rayet mass-loss rate},
  author={St-Louis, N and Moffat, AFJ and Lapointe, L and Efimov, Yu S and Shakhovskoj, NM and Fox, GK and Piirola, V},
  journal={Astrophysical Journal, Part 1 (ISSN 0004-637X), vol. 410, no. 1, p. 342-356.},
  volume={410},
  pages={342--356},
  year={1993}
}

@article{harrington2007spectropolarimetry,
  title={Spectropolarimetry of the H$\alpha$ line in Herbig Ae/Be stars},
  author={Harrington, DM and Kuhn, Jeffrey R},
  journal={The Astrophysical Journal Letters},
  volume={667},
  number={1},
  pages={L89--L92},
  year={2007}
}

@article{schmidt1981spectropolarimetry,
  title={Spectropolarimetry and the physical structure of proto-planetary nebulae},
  author={Schmidt, Gary D and Cohen, Martin},
  journal={Astrophysical Journal, Part 1, vol. 246, June 1, 1981, p. 444-454. Research supported by the University of Minnesota},
  volume={246},
  pages={444--454},
  year={1981}
}

@article{oudmaijer1999halpha,
  title={H$\alpha$ spectropolarimetry of B [e] and Herbig Be stars},
  author={Oudmaijer, Ren{\'e} D and Drew, Janet E},
  journal={Monthly Notices of the Royal Astronomical Society},
  volume={305},
  number={1},
  pages={166--180},
  year={1999},
  publisher={Blackwell Science Ltd 23 Ainslie Place, Edinburgh EH3 6AJ, UK. Telephone~…}
}

@article{eggleton1983approximations,
  title={Approximations to the radii of Roche lobes},
  author={Eggleton, Peter P},
  journal={Astrophysical Journal, Part 1 (ISSN 0004-637X), vol. 268, May 1, 1983, p. 368, 369.},
  volume={268},
  pages={368},
  year={1983}
}

@article{van1993atlas,
  title={An Atlas of High Resolution Line Profiles of Symbiotic Stars-Part One-Coud{\'e} Echelle Spectrometry of Southern Objects and a Classification System of Halpha Line Profiles},
  author={Van Winckel, Hans and Duerbeck, Hilmar W and Schwarz, Hugo E},
  journal={Astronomy and Astrophysics Supplement, Vol. 102, NO. 2/DECI, P. 401, 1993},
  volume={102},
  pages={401},
  year={1993}
}

@article{ivison1994atlas,
  title={An atlas of high resolution line profiles of symbiotic stars. II. Echelle spectroscopy of northern sky objects},
  author={Ivison, RJ and Bode, MF and Meaburn, J},
  journal={Astronomy and Astrophysics Suppl., Vol. 103, p. 201-233 (1994)},
  volume={103},
  pages={201--233},
  year={1994}
}

@article{arrieta2003broad,
  title={Broad H$\alpha$ Wings in Nebulae around Evolved Stars and in Young Planetary Nebulae},
  author={Arrieta, A and Torres-Peimbert, Silvia},
  journal={The Astrophysical Journal Supplement Series},
  volume={147},
  number={1},
  pages={97--102},
  year={2003}
}

@article{chang2015formation,
  title={FORMATION OF RAMAN SCATTERING WINGS AROUND H $\alpha$, H $\beta$, AND PA $\alpha$ IN ACTIVE GALACTIC NUCLEI},
  author={Chang, Seok-Jun and Heo, Jeong-Eun and Mille, Francesco Di and Angeloni, Rodolfo and Palma, Tali and Lee, Hee-Won},
  journal={The Astrophysical Journal},
  volume={814},
  number={2},
  pages={98},
  year={2015},
  publisher={The American Astronomical Society}
}

@article{jung2004centre,
  title={On the centre shifts of Raman-scattered He ii and H$\alpha$ wings in symbiotic stars},
  author={Jung, Yang-Chan and Lee, Hee-Won},
  journal={Monthly Notices of the Royal Astronomical Society},
  volume={350},
  number={2},
  pages={580--586},
  year={2004},
  publisher={Blackwell Science Ltd Oxford, UK}
}

@article{lee2003raman,
  title={Raman-scattered He II $\lambda$6545 line in the symbiotic star V1016 Cygni},
  author={Lee, Hee-Won and Sohn, Young-Jong and Kang, Young Woon and Kim, Ho-Il},
  journal={The Astrophysical Journal},
  volume={598},
  number={1},
  pages={553--559},
  year={2003}
}

@article{lee2012raman,
  title={Raman Scattered He II $\lambda$4332 in the Symbiotic Star V1016 Cygni},
  author={Lee, Hee-Won},
  journal={The Astrophysical Journal},
  volume={750},
  number={2},
  pages={127},
  year={2012},
  publisher={The American Astronomical Society}
}

@article{srivastava2026development,
  title={Development of ProtoPol: a medium resolution echelle spectro-polarimeter for PRL telescopes, Mt. Abu, India—Part I: the design, development, and laboratory characterization},
  author={Srivastava, Mudit K and Maiti, Arijit and Kumar, Vipin and Mistry, Bhaveshkumar and Patel, Ankita and Dixit, Vaibhav and Lad, Kevikumar A},
  journal={Journal of Astronomical Telescopes, Instruments, and Systems},
  volume={12},
  number={2},
  pages={028001--028001},
  year={2026},
  publisher={Society of Photo-Optical Instrumentation Engineers}
}

@article{maiti2026development,
  title={Development of ProtoPol: a medium-resolution echelle spectro-polarimeter for PRL telescopes, Mt. Abu, India—Part II: the data-reduction pipeline, on-sky characterization and performance verification, and first science results},
  author={Maiti, Arijit and Srivastava, Mudit K and Kumar, Vipin and Mistry, Bhaveshkumar and Patel, Ankita and Dixit, Vaibhav and Pandey, Ruchi and Chitroda, Jay},
  journal={Journal of Astronomical Telescopes, Instruments, and Systems},
  volume={12},
  number={2},
  pages={028002--028002},
  year={2026},
  publisher={Society of Photo-Optical Instrumentation Engineers}
}

@article{allen1980unidentified,
  title={On the unidentified bands $\lambda$$\lambda$ 6830, 7088 in symbiotic stars},
  author={Allen, David A},
  journal={Monthly Notices of the Royal Astronomical Society},
  volume={190},
  number={1},
  pages={75--86},
  year={1980},
  publisher={Oxford University Press Oxford, UK}
}

@article{merc2025symbiotic,
  title={Symbiotic stars in the era of modern ground-and space-based surveys},
  author={Merc, Jaroslav},
  journal={Galaxies},
  volume={13},
  number={3},
  pages={49},
  year={2025},
  publisher={MDPI}
}

@article{zhao2025new,
  title={New Symbiotic Stars or Candidates in LAMOST Low-resolution Spectra},
  author={Zhao, Yabing and Guo, Sufen and L{\"u}, Guoliang and Zhu, Chunhua and Li, Jiao and Shi, Jianrong},
  journal={The Astrophysical Journal},
  volume={995},
  number={1},
  pages={14},
  year={2025},
  publisher={The American Astronomical Society}
}

@article{chen2025new,
  title={New Symbiotic Stars from LAMOST DR10 Spectra and Multiband Photometry},
  author={Chen, Jing and Wang, Liang and Li, Yin-Bi and Ma, Xiao-Xiao and Luo, A-Li and Zhang, Zi-Chong and Ding, Ming-Yi and Zhang, Kai},
  journal={The Astrophysical Journal},
  volume={987},
  number={2},
  pages={147},
  year={2025},
  publisher={The American Astronomical Society}
}

@article{kafatos1991ultraviolet,
  title={Ultraviolet and optical spectroscopy of the R Aquarii symmetrical jet},
  author={Kafatos, Menas},
  journal={The Astrophysical Journal},
  year={1991}
}

@article{seaquist1984radio,
  title={A radio survey of symbiotic stars},
  author={Seaquist, ER and Taylor, AR and Button, S},
  journal={Astrophysical Journal, Part 1 (ISSN 0004-637X), vol. 284, Sept. 1, 1984, p. 202-210.},
  volume={284},
  pages={202--210},
  year={1984}
}

@article{ball2025symbiotic,
  title={Symbiotic star candidates in Gaia Data Release 3},
  author={Ball, Samantha E and Bromley, Benjamin C and Kenyon, Scott J},
  journal={arXiv preprint arXiv:2506.20505},
  year={2025}
}

@article{akras2026discovery,
  title={Discovery of thirteen new symbiotic stars in Gaia DR3},
  author={Akras, Stavros and Karagiannis, Anastasis and Charalampopoulos, Giorgos and Gavras, Panagiotis and Guti{\'e}rrez-Soto, Luis A},
  journal={Monthly Notices of the Royal Astronomical Society},
  pages={stag105},
  year={2026},
  publisher={Oxford University Press}
}

@article{mukai2016lyncis,
  title={SU Lyncis, a hard X-ray bright M giant: Clues point to a large hidden population of symbiotic stars},
  author={Mukai, K and Luna, Gerardo Juan Manuel and Cusumano, G and Segreto, A and Munari, U and Sokoloski, JL and Lucy, AB and Nelson, T and Nu{\~n}ez, Natalia Edith},
  journal={Monthly Notices of the Royal Astronomical Society: Letters},
  volume={461},
  number={1},
  pages={L1--L5},
  year={2016},
  publisher={The Royal Astronomical Society}
}

@article{kumar2020uv,
  title={UV spectroscopy confirms SU Lyn to be a symbiotic star},
  author={Kumar, Vipin and Srivastava, Mudit K and Banerjee, Dipankar PK and Joshi, Vishal},
  journal={Monthly Notices of the Royal Astronomical Society: Letters},
  volume={500},
  number={1},
  pages={L12--L16},
  year={2020},
  publisher={Oxford University Press}
}

@article{ilkiewicz2022lyn,
  title={SU Lyn-a transient symbiotic star},
  author={I{\l}kiewicz, Krystian and Miko{\l}ajewska, Joanna and Scaringi, Simone and Teyssier, Francois and Stoyanov, Kiril A and Fratta, Matteo},
  journal={Monthly Notices of the Royal Astronomical Society},
  volume={510},
  number={2},
  pages={2707--2717},
  year={2022},
  publisher={Oxford University Press}
}

@article{burmeister2009spectroscopy,
  title={Spectroscopy of the symbiotic binary CH Cygni from 1996 to 2007},
  author={Burmeister, M and Leedj{\"a}rv, L},
  journal={Astronomy \& Astrophysics},
  volume={504},
  number={1},
  pages={171--180},
  year={2009},
  publisher={EDP Sciences}
}
\bibliographystyle{aasjournalv7}



\end{document}